\documentclass[3p, preprint]{elsarticle}

\usepackage{array}
\usepackage{graphicx}
\usepackage{listings}
\usepackage{caption}
\usepackage{adjustbox}
\usepackage{algorithm}
\usepackage{hyperref}
\usepackage{setspace}
\usepackage{tabularx}
\usepackage{color}
\usepackage{float}
\usepackage{wrapfig}
\usepackage{booktabs}
\usepackage{graphicx,subcaption}
\usepackage{amssymb}
\usepackage{pifont}

\usepackage{rotating}
\usepackage{afterpage}
\usepackage{longtable}
\usepackage{lscape} 

\usepackage{pgfplots}
\pgfplotsset{compat=newest}
\usetikzlibrary{patterns}

\usepackage{amsmath,amssymb,amsfonts}
\usepackage{bm}
\usepackage[inline]{enumitem}

\usepackage{algpseudocode}
\usepackage{adjustbox}
\usepackage{verbatim}

\usepackage{booktabs, multirow} 
\usepackage{soul}
\usepackage{longtable}

\newboolean{showcomments}
\setboolean{showcomments}{false}         

\ifthenelse{\boolean{showcomments}}
  {\newcommand{\nb}[2]{
  \fbox{\bfseries\sffamily\scriptsize#1}
     {\sf\small$\blacktriangleright$\textit{\textcolor{red}{#2}}$\blacktriangleleft$}
   }
  }
  {\newcommand{\nb}[2]{}
   
  }

\newcommand{\methodname}[0]{PARIOT}
\newcommand{\toolname}[0]{PARIOTIC}
\newcommand{\moduleone}[0]{CLB Injector}
\newcommand{\moduletwo}[0]{CLB Protector}

\definecolor{dkgreen}{rgb}{0,0.6,0}
\definecolor{gray}{rgb}{0.5,0.5,0.5}
\definecolor{mauve}{rgb}{0.58,0,0.82}

\begin{document}

\title{PARIOT: Anti-Repackaging for IoT Firmware Integrity}  

\author[1]{Luca Verderame \corref{cor1}}
\ead{luca.verderame@dibris.unige.it} 

\author[1]{Antonio Ruggia}\ead{antonio.ruggia@dibris.unige.it} 

\author[1]{Alessio Merlo}\ead{alessio@dibris.unige.it} 

\address[1]{DIBRIS - University of Genoa, Via Dodecaneso, 35, I-16146, Genoa, Italy.}
\cortext[cor1]{Corresponding author}


\begin{abstract}
IoT repackaging refers to an attack devoted to tampering with a legitimate firmware package by modifying its content (e.g., injecting some malicious code) and re-distributing it in the wild. In such a scenario, the firmware delivery and update processes are central to ensuring firmware integrity. 

Unfortunately, several existing solutions lack proper integrity verification, exposing firmware to repackaging attacks. If this is not the case, they still require an external trust anchor (e.g., signing keys or secure storage technologies), which could limit their adoption in resource-constrained environments. 
In addition, state-of-the-art frameworks do not cope with the entire firmware production and delivery process, thereby failing to protect the content generated by the firmware producers through the whole supply chain.

 To mitigate such a problem, in this paper, we introduce \methodname{}, a novel self-protecting scheme for IoT that injects integrity checks, called anti-tampering (AT) controls, directly into the firmware.
The AT controls enable the runtime detection of repackaging attempts without needing signing keys, internet connection, secure storage technologies, or external trusted parties. 
\methodname{} can be adopted on top of existing state-of-the-art solutions ensuring the widest compatibility with current IoT ecosystems and update frameworks.
Also, we have implemented this scheme into \toolname{}, a prototype to protect C/C++ IoT firmware automatically. The evaluation phase of 50 real-world firmware samples demonstrated the proposed methodology's feasibility and robustness against practical repackaging attacks without altering the firmware behavior or severe overheads.
\end{abstract}

\maketitle

\section{Introduction}

The Internet of Things (IoT) paradigm enables the growth of low-cost embedded devices with network connectivity and real-time capabilities that are now used in many verticals, from logistics to precision farming and smart homes. 
Each IoT device is equipped with firmware, i.e., a bundle that contains all the software needed to ensure the functioning of the device hardware. Typically, the firmware comprises a fully-fledged IoT operating system (like RIOT \cite{baccelli2013riot} or Contiki \cite{dunkels2004contiki}) and at least an application that holds the core functionalities of the thing.

During the building phase, the device manufacturer equips the IoT device with the first version of the firmware. However, the functionalities an IoT device requires at deployment time will likely change. To this aim, the firmware will need frequent updates for several reasons: to offer additional functionalities, support new communication protocols, and patch software bugs (including security vulnerabilities).

As firmware has a central role in the life cycle of an IoT device, its security has raised serious concerns from the scientific and industrial community. To this aim, several works were proposed to evaluate the security of the firmware bundle (e.g., \cite{david2018firmup}, \cite{costin2014large} or \cite{costin2016automated}), enforce the update mechanisms (e.g., \cite{dejon2019automated}, \cite{cui2013firmware} and \cite{langiu2019upkit}), or patch existing firmware bundles to cope with end-of-life, vulnerable images. 
 (e.g., \cite{carrillo2022hale}, \cite{maroof2022irecover}, and \cite{christensen2020decaf}).

In particular, the integrity of the firmware delivered through an update process represents a major security threat, as witnessed by many real-life examples. For instance, the PsycoB0t \cite{pycobot} was the first router botnet that altered the firmware of approximately 85,000 home routers and resulted in large-scale denial of service attacks. Also, the Zigbee Worm \cite{ronen2017iot} triggered a chain reaction of infections, initialized by a single compromised IoT device (light bulb), using a malicious firmware update image.

An attacker can retrieve the firmware differently, such as obtaining it from the vendor's website or community forums, sniffing the OTA update mechanism, or dumping it directly from the device \cite{gupta2019iot}.
Once the original firmware is obtained, 
the attacker can analyze it through reverse engineering techniques to extract sensitive information such as encryption keys, hard-coded credentials, or internal URLs.
Thanks to such knowledge, an attacker can craft a modified version of the firmware and try to re-distribute it in the wild as if it was the original one \cite{mtetwa2019secure}.

This type of attack called \emph{repackaging}, is well-known in the mobile ecosystem, where attackers alter and re-distribute thousands of Android and iOS applications \cite{MERLO2021102181}.
Unfortunately, such a security threat is barely considered in the IoT ecosystem, especially in low-end devices where resource constraints limit the applicability of state-of-the-art mitigation techniques such as remote attestation or signature verification.  

Furthermore, many of the existing solutions for low-end IoT devices focus only on some parts of the delivery process (e.g., from the update server to the device) or do not perform a proper verification of the downloaded firmware and hence cannot ensure its integrity. 
For instance, Sparrow \cite{sparrow} (used by Contiki) only verifies the CRC of the image to detect errors during transmissions.

On the other hand, recent firmware update solutions like SUIT \cite{suit} or UpKit \cite{langiu2019upkit} need to have an additional trust anchor (e.g., a signing certificate on the IoT device) or dedicated hardware, like a Trusted Execution Environment \cite{asokan2018assured}, to allow verifying the integrity of the image.
Nevertheless, the impairment of the supply chain or the delivery mechanism (e.g., the update server or the companion mobile app) could allow an attacker to inject a crafted firmware into the delivery pipeline, such as the firmware modification attack reported on commercial fitness trackers \cite{shim2017case}. 

In this work, we investigate the impact of repackaging attacks on IoT firmware, thereby discussing security threats that harm its integrity. Also, we systematically review the integrity protection mechanisms used by state-of-the-art solutions for IoT firmware updates, unveiling their need to rely on signing keys, internet connection, or trusted external entities to cope with the integrity of firmware bundles.

Then, we present \methodname{}, a novel self-integrity protection mechanism for IoT firmware that automatically spots altered firmware images without any prerequisite to ensure the integrity of the firmware image or modification to existing firmware delivery infrastructure.

Briefly, \methodname{} focuses on inserting encrypted detection nodes (called \textsl{Cryptographically Obfuscated Logic Bombs} \cite{10.1145/3168820}) that embed integrity checks on the content of the firmware.
These checks are known as \textsl{anti-tampering} (AT) controls.
The detection nodes are triggered during the execution of the firmware, and if some tampering is detected, the firmware is usually forced to crash. The rationale is to discourage the attacker from repackaging if the likelihood of building a working repackaged firmware is low. 

\methodname{} aims at applying the protection scheme on the source code during the building pipeline to ensure the widest compatibility with state-of-the-art software update methods (e.g., SUIT) and at minimizing the use of invasive procedures (e.g., binary rewriting \cite{wenzl2019hack}) that may harm/brick the firmware image and, ultimately, the IoT device.

To experimentally evaluate the feasibility of \methodname{}, we implemented it in a tool for C/C++ firmware (i.e., \toolname{}) that is compatible with existing firmware update solutions. We tested \toolname{} in a RIOT-based IoT ecosystem with the SUIT update framework and 50 real-world firmware samples. 
The tool - publicly available on GitHub \cite{patriot} - achieved a 90\% of success rate and required - on average - only 64.2 seconds per firmware to introduce the protections. 
We performed the runtime evaluation on an \textit{iotlab-m3 board} hosted by the FIT IoT-LAB testbed~\cite{adjih2015fit} to verify the resource consumption overhead (in terms of current, voltage, and power) introduced by the protection on real IoT devices.
Moreover, we assessed the reliability of \methodname{} by testing the repackaging detection capabilities of the protected firmware at runtime.
The preliminary results showed that our solution ensures high compatibility with existing firmware generation and delivery processes, a low resource usage overhead, and a high detection rate of repackaging attacks.

\textbf{Structure of the paper.} In the rest of the paper, we first introduce the firmware production and delivery process (Section \ref{sec:delivery}). 
Then, we focus on the integrity threats concerning the previous steps (Section \ref{sec:threat}). 
Section \ref{sec:background} presents the concept of firmware repackaging in the IoT ecosystem and discusses state-of-the-art anti-repackaging techniques. In contrast, Section \ref{sec:related} highlights the limitations of state-of-the-art approaches.

In Section \ref{sec:methodology}, we describe the \methodname{} protection scheme, its distinguishing features, and its runtime behavior. 
Moreover, we provide an implementation of the methodology (\toolname{}) in Section \ref{sec:implementation}. 
Section \ref{sec:experiments} analyzes the results obtained by applying \toolname{} on 50 real-world firmware in a RIOT-based IoT ecosystem.
Finally, in Section \ref{sec:conclusion}, we conclude the paper by summing up the main takeaways and putting forward some considerations for future works.

\section{Firmware Production and Delivery Process}\label{sec:delivery}

\begin{figure*}[!ht]
  \includegraphics[width=\textwidth]{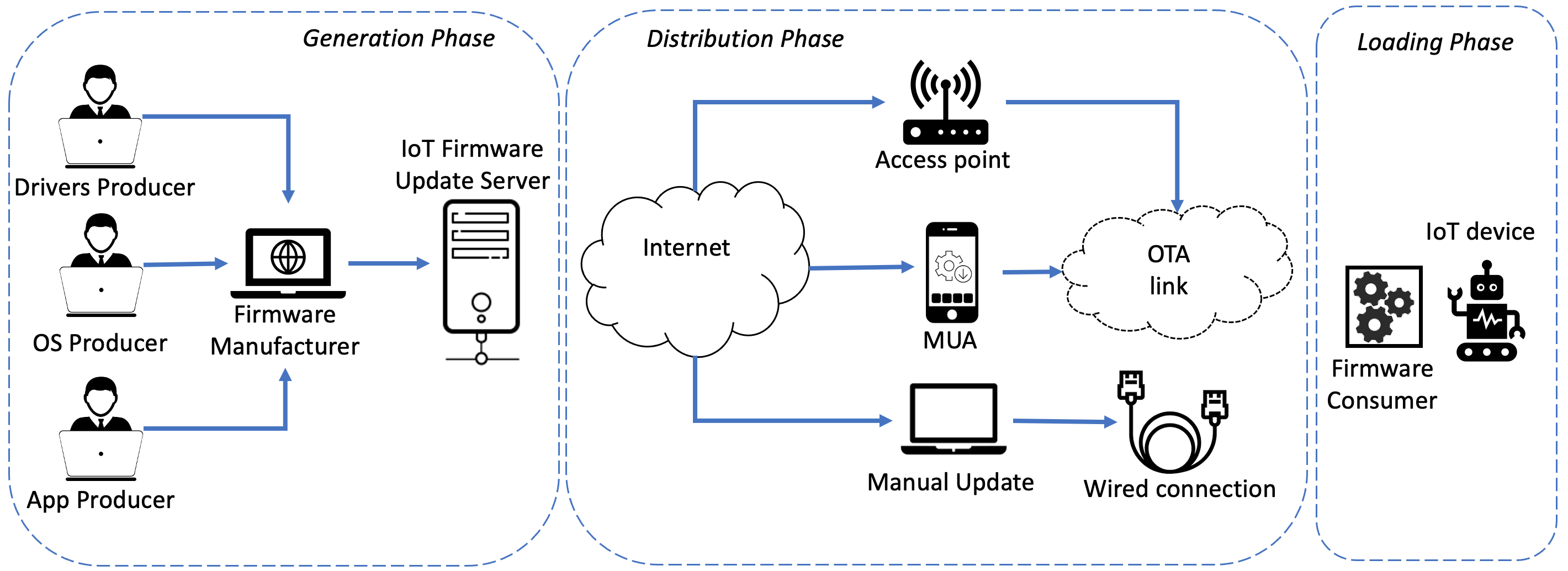}
  \caption{Firmware Production and Delivery Process.}
  \label{fig:firmware-delivery}
\end{figure*}

Figure \ref{fig:firmware-delivery} summarizes a typical firmware production and delivery process. The first steps are devoted to the production of the firmware (\emph{generation phase}), i.e., the building of the software bundle containing all the software that ensures the functioning of the IoT device. 

The firmware supply chain - even for a relatively simple, single-processor device - consists of many software providers, including chip and tool vendors and companies that provide different software components. In the case of Figure \ref{fig:firmware-delivery}, the software supply chain comprises three actors providing the OS, the device drivers, and the application, which delivers device core functionalities, respectively. 

The different pieces of software are then composed by the \textit{Firmware Manufacturer} (FM), which generates the firmware image and some metadata information, like a digitally-signed manifest file, used to evaluate the success of the delivery phase.
At the end of the generation phase, the firmware is released on a centralized repository (e.g., an IoT Firmware Update Server).

The firmware delivery process (\emph{distribution phase}) can occur either \emph{manually} or by employing an \emph{automated firmware update} process.
In the first case, users get the firmware from the firmware repository and distribute it to the IoT devices by using over-the-air (OTA) technologies (e.g., Bluetooth LE or Wi-Fi) or physical interfaces (e.g., UART or USB ports).
For such a task, the user may also rely on a \textit{Mobile Update Agent} (MUA), i.e., a companion mobile app (e.g., Samsung SmartThing\footnote{\url{https://www.samsung.com/it/apps/smartthings/}}) that acts as a gateway between the update server and the IoT device. 
After receiving an update notification, the MUA downloads the software update and verifies its integrity, and - once the verification succeeds - it sends the update to the \textit{Firmware Consumer} using low-power radio technologies.
As an example of integrity control, the MUA can verify the consistency between the digitally-signed manifest file and the software package. 

In the automated distribution phase, instead, the firmware is distributed to the IoT devices by using client-server architectures (e.g., Software Updates for Internet of Things -- SUIT) or distributed solutions (e.g., \cite{choi2020blockchain}). 
If this is the case, the IoT Firmware update server interacts through an access point with the device.

The last step in the firmware update is the \emph{loading} phase. 
An agent placed on the device (i.e., the \textit{Firmware Consumer} - FC) receives the software bundle and the metadata and copies the updated image in the correct memory address to proceed with the installation. Such a step may involve a further verification of the correctness of the received data, e.g., through a hash check or signature verification.

\section{Threats to Firmware Integrity}\label{sec:threat}

Security threats involving the integrity of firmware bundles can occur in all three stages of the production and delivery process. 
This section aims to provide information about the threats targeting the integrity of the firmware bundle, the phase and the entities that are affected, and the mitigation techniques and the requirements that enable to cope with those threats.

From our analysis, we identified eight distinct security threats harming the integrity of the firmware updates, with some affecting more than a phase of the firmware production and delivery process. The rest of this section provides a brief description of each of them.
Table \ref{tab:threats} reports the list of all security threats, which can be uniquely identified by a \textit{Thread ID} (first column).
To do so, we exploited the Spoofing, Tampering, Repudiation, Information Disclosure, Denial of Service, and Elevation of Privilege (STRIDE) approach \cite{khan2017stride} and integrated the contributions of state-of-the-art research in the field \cite{david2018firmup,gupta2019iot,suit,lanigan2006sluice}. 

\paragraph{\textbf{Modification of firmware before signing}} If an attacker can alter the firmware bundle before it is signed (e.g., by modifying the code of one of its components) during the generation phase, she can perform all the same actions as the firmware manufacturer. 
This allows the attacker to deploy firmware updates to devices that trust the FM. 
For example, the attacker that deploys malware in the building environment of one of the Producers or the FM can inject code into any binary referenced by the bundle, or she can replace the referenced binary (digest) and URI with the attacker’s ones. 
Possible mitigation techniques are a validation process of the firmware bundle (e.g., by enforcing vulnerability assessment and penetration testing activities) and using air-gapped building environments that are protected from external interference. 

\paragraph{\textbf{Overriding critical metadata elements}}
An authorized actor - but not the FM - uses an override mechanism during the generation phase to change an information element in the metadata signed by the FM. 
For example, if the authorized actor overrides the digest and URI of the payload in the manifest file, she can replace the entire payload with another - properly crafted -one. 
To mitigate this threat, the firmware update process should enforce mandatory access control-like mechanism by using access control lists with per-actor rights enforcement of the FM and the different producers \cite{suit}.

\paragraph{\textbf{Compromission of the intermediate agents}}
If an attacker succeeds in compromising an agent in the distribution phase, then she can inject a malicious/modified firmware bundle into the distribution chain. 
For instance, a malicious actor can compromise the MUA to replace the original firmware with a modified version after the firmware verification phase, thus installing a modified firmware on the IoT device: Sch{\"u}ll et al. \cite{schull2016data} showed a new firmware modification attack against a fitness tracker, where an adversary manipulated plain HTTP traffic and TLS proxy between an original gateway and the update server.
This threat can be mitigated by providing a secure distribution environment (e.g., a security-hardened MUA) and implementing a signature verification process on the FC. 

\paragraph{\textbf{Traffic interception}}
In such a scenario, an attacker intercepts all traffic to and from a device with the ability to monitor or modify any data sent to or received from the device \cite{mtetwa2019secure}. This capability allows an attacker to alter or drop a valid firmware bundle or its associated metadata during the distribution phase.

This threat can be mitigated by enforcing a secure transmission protocol and/or encrypting the exchanged data \cite{nguyen2015survey}. 

\paragraph{\textbf{Image replacement on the device}}
In this scenario, the attacker replaces a newly downloaded firmware after the device finishes verifying its metadata (e.g., it executes integrity checks on the manifest file), fooling the device into executing the attacker's image.
This attack likely requires physical access to the device; however, it can be carried with another threat that allows remote execution. 
Common mitigation techniques consist of adopting a verification mechanism of the firmware bundle (e.g., signature/digest verification) on the FC and storing the bundle in immutable/protected memory \cite{suit}. 

\paragraph{\textbf{Modification of metadata between authentication and use}}
If an attacker can modify the metadata information after it is authenticated (Time Of Check) but before it is used (Time Of Use) \cite{mtetwa2019secure}. The attacker can replace any content whatsoever. For instance, she can replace the URI of the firmware bundle in the update manifest file after it is validated by the MUA, causing the FC to download and install a repackaged firmware. 
This threat can be mitigated by enforcing the verification of the metadata on the FC and not on intermediate agents (e.g., the MUA). 

\paragraph{\textbf{Exposure of signing keys}}
If an attacker obtains a key or even indirect access to a key, then she can perform the same actions as the legitimate user. 
In the worst case, if the key retrieved by the attacker is considered trusted by the firmware update chain, the attacker can perform firmware updates as though they were the legitimate owner of the key.
For example, if the attacker can obtain the Firmware Manufacturer's signing key, she can generate malicious firmware updates and deliver them through the distribution framework.
This threat can be mitigated by storing the signing keys in a protected/separated storage, implementing a key rotation mechanism, or using air-gapped devices to execute the signing process.  

\paragraph{\textbf{Unauthenticated images}}
In the case the IoT device does not verify the image, an attacker can install a custom firmware on a device by, for example, manipulating either the payload or the metadata gaining complete control of the device. 
This attack can be prevented by introducing digitally signed metadata that can be verified by the FC \cite{suit}. 

\begin{table}[!htp]\centering
\tiny
\begin{tabular}{llll}\toprule
\textbf{ID} &\textbf{Name} &\textbf{Phase} &\textbf{Involved Entities} \\\midrule
IMG.MODIFICATION &Modification of firmware prior to signing &Generation &Producers, FM \\
META.OVERRIDE &Overriding critical metadata elements &Generation &Producers, FM \\
DIS.AGCOMPR &Compromission of the intermediate agents &Distribution &Access Point, MUA, IoT Client \\
DIS.MITM &Traffic interception &Distribution &Network, OTA links \\
IMG.REPLACE &Image replacement on the device &Distribution, Loading &MUA, IoT device \\
META.TOCTOU &Modification of metadata between authentication and use &Distribution, Loading &MUA, IoT device \\
KEY.EXPOSURE &Exposure of signing keys &Generation, Loading &FM, IoT device \\
IMG.NON\_AUTH &Unauthenticated images &Loading &IoT device \\
\bottomrule
\end{tabular}
\caption{List of threats affecting the integrity of firmware during the Firmware Production and Delivery Process.}\label{tab:threats}
\end{table}

\section{Firmware Repackaging: Background, Attacker Model \& Countermeasures}
\label{sec:background}

The goal of a firmware repackaging attack \cite{panchal2018security} is to tamper with legitimate firmware and redistribute the modified - i.e., \textit{repackaged} - version to IoT devices to perform further attacks.
Repackaging attacks on IoT firmware are motivated by at least one of the following reasons:
\begin{itemize}
    \item \textbf{Unauthorized access/usage.} The attacker modifies the firmware to bypass the predefined software access control privileges, e.g., to gain access to privileged functionalities or classified data.
    \item \textbf{Unlicensed clones.} The attacker's goal is to reuse the crucial firmware processes in some other programs. To this aim, the attacker can extract and partially reuse part of the firmware image to craft a clone of the original version \cite{al2015framework}.
    \item \textbf{Malware injection.} By injecting malicious code, the attacker aims to breach the firmware integrity, thereby illegally altering the firmware behavior. In such a case, compromised firmware can disrupt the trustworthiness of an IoT device \cite{zdnet2016}.
    \item \textbf{Disrupting the system availability.} The attacker aims to reduce the system availability, injecting code in the firmware image able to cause system halting (like DoS Attack) or significant delays in the regular operation of an IoT device.
\end{itemize}

\begin{figure}[h]
  \centering
  \includegraphics[width=0.76\linewidth]{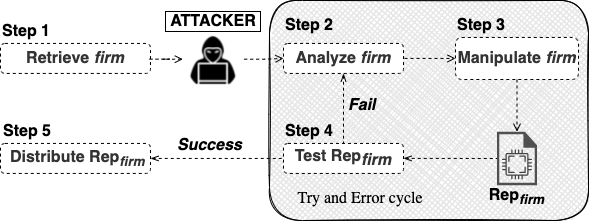}
  \caption{Steps of a \emph{firmware repackaging} attack.}
  \label{fig:repack-steps}
\end{figure}

To carry out a repackaging attack, a malicious actor jointly exploits the security threats discussed in Section \ref{sec:threat}.
Figure \ref{fig:repack-steps} illustrates the steps involved in the firmware repackaging attack.

The first step is retrieving the original firmware image ($firm$ in Figure \ref{fig:repack-steps}). 
Attackers can get hold of the firmware by: \emph{i)} dumping it using the physical/remote access interfaces of an IoT device or an intermediate agent \cite{vasile2018breaking} (DIS.AGCOMPR), \emph{ii)} sniffing the package during an over the air (OTA)  update \cite{gupta2019iot} (DIS.MITM), or \emph{iii)} downloading it from the vendor's website,  support, and community forums, or public repositories.

Then, the attacker analyzes the target firmware (Step 2) using static and dynamic analysis techniques to inspect its behavior and extract sensitive information such as encryption keys, hard-coded credentials, or sensitive URLs. 
Notable tools to perform such analysis include Binwalk \cite{binwalk}, Firmware Analysis Comparison Toolkit (FACT ) \cite{fact}, Firmware Modification Kit \cite{fmk}, and Firmwalker \cite{firmwalker}. 
This is a crucial step that allows an attacker to thoroughly understand the firmware bundle, exposing the firmware producer to security and privacy issues.
The extracted knowledge will be used in Step 3 to manipulate the original image to inject custom code (e.g., malware), modify the existing binary image to alter the legitimate behavior (IMG.MODIFICATION) or override critical metadata elements (META.OVERRIDE). In this step, the attacker can exploit binary rewriting techniques \cite{wenzl2019hack} to produce a crafted version of the firmware, i.e., $REP_{firm}$.

Finally, the attacker tests whether $REP_{firm}$ works properly (Step 4). If this is the case, in Step 5, the attacker redistributes $REP_{firm}$ in the delivery pipeline (e.g., DIS.AGCOMPR) or installs it directly on a target device (e.g., IMG.NON\_AUTH, or META.TOCTOU).
Otherwise, the attacker further analyzes and modifies the original $firm$ (back to Step 2). Steps 2 to 4 are also known as the \textsl{try and error cycle} that the attacker must keep executing until she gets a working repackaged firmware.

\subsection{Anti-Repackaging Techniques}

\begin{figure}[h]
  \centering
  \includegraphics[width=0.76\linewidth]{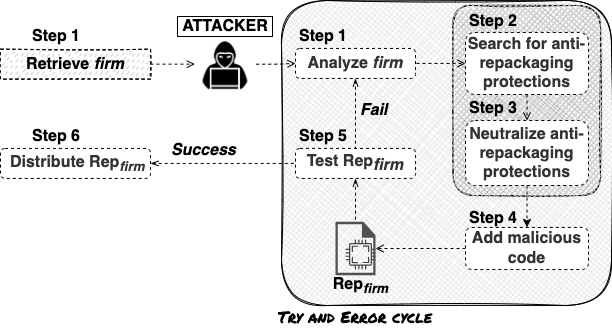}
  \caption{Steps of a \emph{firmware repackaging} attack for anti-repackaging protection.}
  \label{fig:repack-steps-at}
\end{figure}

Anti-repackaging aims to protect software from being successfully repackaged. 
In the context of the IoT update process, anti-repackaging protects the integrity of the entire firmware to ensure that the IoT device will download, install, and execute the expected update.
From the attacker's side, the activities to mount a repackaging attack will now include two additional steps (i.e., Steps 2 and 3 of Figure \ref{fig:repack-steps-at}) related to detecting and disabling repackaging protection techniques.

To this aim, an ideal anti-repackaging solution never lets the attacker obtain working repackaged firmware (i.e., moving out from the \textsl{try and error cycle}). A reliable anti-repackaging solution makes the repackaging non-cost-effective, i.e., it requires so much time to be disabled that the attacker gives up on repackaging the target.

Depending on the type of techniques enforced on the pipeline, we can distinguish between two categories:

\paragraph{\textbf{External Anti-repackaging}} 
This category includes all the techniques relying on an external agent to execute anti-tampering checks. Examples are update agents \cite{langiu2019upkit}, trusted servers \cite{asokan2018assured}, or even blockchains \cite{lee2017blockchain}. 
For example, the authors of \cite{perito2010secure} propose a Secure Code Update By Attestation in sensor networks (SCUBA), which can be used to repair a compromised sensor through firmware updates. SCUBA utilizes an authentication mechanism and software-based attestation to identify memory regions infected by malware and transmits the repair update to replace these regions. However, the attestation technique based on self-checksumming code heavily relies on consistent timing characteristics of the measurement process and an optimal checksum function. Due to these assumptions, SCUBA is not a suitable approach for IoT settings.

To achieve the goal of protection, these techniques face the challenge of creating a communication channel between the firmware and the external authority to perform the check. However, such a prerequisite may not be feasible in real-world scenarios of IoT ecosystems with low connectivity or limited computational capabilities. 
In addition, this channel has to be protected from repackaging, exposing it as a single point of failure.

\paragraph{\textbf{Internal Anti-repackaging}}
This category of techniques aims to protect software from being successfully repackaged by adding some protection code - called \textsl{detection nodes} - in the source code before building the firmware image and delivering it to the distribution phase.

The idea of detection nodes has been put forward in \cite{7579771} and it refers to a self-protecting mechanism made by a piece of code inserted into the original software, which carries out integrity checks - called \textsl{anti-tampering} controls (e.g., signature check, package name check) - when executed at runtime.
More specifically, anti-tampering checks compare the signature of a specific part of the firmware with a value pre-computed during the building of the original bundle; if such values differ, then a repackaging is detected, and the detection node usually leads the firmware to fail, thereby frustrating the repackaging effort.

To protect detection nodes, anti-repackaging solutions hide them into the so-called \textsl{logic bombs} \cite{10.1145/3168820}, which has been originally conceived in the malware world to hide malicious payloads \cite{sharif2008impeding}.
A logic bomb is a piece of code that is executed when specific conditions are met. While logic bombs are widely used by malware to introduce and trigger malicious conditions inside apparently unarmed code \cite{7546513, Brumley2008}, this technique can also include tampering detection code (namely, AT checks) inside the activated bomb. 
In a nutshell, anti-tampering checks are self-protecting functions that aim to detect modifications in a piece of software.
To this goal, AT checks may verify at runtime some relevant information of the code or the whole executable file against precomputed values. 
Several methods exist to detect tampering in executable files, as highlighted in \cite{AHMADVAND2019413} and \cite{eldefrawy2012smart}.

To hide the behavior of a logic bomb, researchers introduced the concept of the Cryptographically Obfuscated Logic Bomb (hereafter, CLB).
At build time, the content of a logic bomb is replaced by an encrypted version, which is only decrypted at runtime.
In a typical scenario, to perform encryption, the developer should include (and thus reveal) the decryption key(s) in the executable.
To avoid revealing the key, the authors in~\cite{10.1145/3168820} proposed a novel form of CLB that exploits the executable's logic to hide the key value.
In particular, the use of a CLB (Listing~\ref{lst:clb}) consists in embedding a logic bomb in a \textit{qualified condition} (QC) and replacing it with an encrypted form.
A QC is a branch (if statement) containing an equality check where one of the operands is a constant value (e.g., \texttt{X == const}).
Thus, since inside a QC, the variable (i.e., \texttt{X}) is \textit{always} equal to the constant value (i.e., \texttt{const}), \texttt{X} can be used as a decryption key for the content of the branch.
It is worth noting that the CLB is self-contained and does not require an external or internal thrust anchor: a developer should not include and expose the decryption key(s), which is automatically computed by the program at runtime.

To create this kind of CLB, the original condition is transformed into a new one where the pre-computed hash value of the constant (i.e., \texttt{Hconst}) is compared with the result of the hash function applied to variable \texttt{X} plus some \texttt{salt}.
Besides, the original content of the qualifying condition is 
encrypted using the \texttt{const} value as the encryption key (i.e., \texttt{encrypted\_content}). 
If the triggering condition is met, then the \texttt{X} value is used to decrypt the \texttt{encrypted\_content} and, thus, launch the bomb.

\begin{lstlisting}[caption={Example of Cryptographically Obfuscated Logic Bomb (CLB).}, label={lst:clb}]
if(H(X,salt) == Hconst){
    body = decrypt(encrypted_content, X);
    execute(body);
}
\end{lstlisting}

Logic bombs rely on the information asymmetry between the developer and the attacker, i.e., since the attacker has partial knowledge of the software behaviors, it is unlikely that she can correctly guess the key value used to encrypt the code. The protection of the CLB is granted by the one-way property of cryptographic hash functions, which makes it hard for the attacker to retrieve the original \texttt{const} value (i.e., the decryption key) from its hash.

\section{Related Works}
\label{sec:related}

Operating systems designed explicitly for constrained IoT devices (e.g., TinyOS [30] and Contiki [31]) often embed or can be extended with over-the-air reprogramming capabilities.
Still, many existing solutions focus only on a portion of the update process or do not properly verify the downloaded firmware and hence cannot ensure its integrity.
This indeed results in OSes without an update system (e.g., NuttX \cite{nuttx}) or with an incomplete one. For instance, Sparrow \cite{sparrow} (used by Contiki) and Deluge \cite{hui2004dynamic} (used by TinyOS) only verify the CRC to ensure the integrity of the firmware during transmissions, leading to the possibility of abusing the lack of controls by an attacker (e.g., through IMG.MODIFICATION, META.TOCTOU, and IMG.NON\_AUTH threats). 

To this aim, the research community has been working on the definition of secure IoT update processes \cite{arakadakis2020firmware}, \cite{el2022secure}.
For instance, \cite{hyun2008seluge}, \cite{lanigan2006sluice}, and \cite{dutta2006securing} proposed secure extensions for Deluge that provide integrity assurance for the firmware image and resilience against DoS attacks that specifically target firmware dissemination protocols.

Other solutions, like UpKit \cite{langiu2019upkit} or ASSURED \cite{asokan2018assured}, put forward scalable and lightweight approaches able to perform software updates with end-to-end protection across different OSes and hardware platforms.
Also, the authors in \cite{zandberg2019secure}  propose a secure firmware update mechanism for constrained IoT devices based on open standards such as CoAP, LwM2M, and SUIT.

Finally, several works \cite{lee2017blockchain,hu2019autonomous,yohan2018over} exploited the blockchain technology to verify the authenticity and integrity of a firmware version and to distribute a specific version of firmware binary to the connected nodes in the blockchain network. Notable examples include Firmware-Over-The-Blockchain (FOTB) \cite{yohan2020fotb} and CHAINIAC \cite{nikitin2017chainiac} that exploit Bitcoin and Ethereum blockchain, respectively.

In our work, we reviewed the main state-of-the-art firmware update solutions to identify the adopted integrity protection mechanisms, the mandatory requirements to ensure their proper functioning and the involved entities of the Firmware Production and Delivery Process. Table~\ref{tab:sota} reports the results of the analysis.

Unfortunately, most of the existing proposals build the integrity and authenticity of updates on one or more signing keys, which are prone to loss \cite{dlink15}, theft \cite{freebsd12}, or misuse \cite{hynday22} (i.e., KEY.EXPOSURE threat).
Proper protection for signing keys to defend against such single points of failure is a top priority but requires secure storage technologies such as hardware security modules \cite{asokan2018assured}).
It is worth emphasizing that revoking and renewing signing keys (e.g., in reaction to a compromise) and informing all their clients about these changes is usually cumbersome.

Also, some existing techniques face the challenge of creating a communication channel between the IoT device and the external authority (e.g., the blockchain or a tamper-proof server) to perform the check (e.g., DIS.MITM, or IMG.REPLACE threats). However, such a prerequisite may not be feasible in real-world scenarios of IoT ecosystems with low connectivity or limited computational capabilities.

To overcome such limitations, in this paper, we will present \methodname{}. This first solution ensures the integrity of the firmware update by relying on a self-protection mechanism that does not require signing keys, internet connection, secure storage technologies, or external trusted parties.
In addition, our methodology is agnostic w.r.t. the firmware delivery model and, thus, can be adopted on top of existing solutions providing an extra layer of security with negligible overhead.

\begin{scriptsize}
\begin{longtable}{>{\color{black}}l >{\color{black}}l >{\color{black}}l >{\color{black}}l >{\color{black}}l}
\toprule
\textbf{Work} &\textbf{Year} &\textbf{Integrity Protection Mechanism} &\textbf{Requirements} \\\midrule
\cite{lanigan2006sluice} &2006 &\begin{tabular}{@{}l@{}}ECDSA signature verification of each \\transmitted block\end{tabular} &\begin{tabular}{@{}l@{}}- Pub Keys (IoT Devices) \\ - Secure Storage Technology for Private Key (FM)\end{tabular} \\
\midrule
\cite{dutta2006securing} &2006 &\begin{tabular}{@{}l@{}}RSA signature verification of each transmitted \\block\end{tabular} &\begin{tabular}{@{}l@{}}- Pub Keys (IoT Devices) \\ - Secure Storage Technology for Private Key (FM)\end{tabular} \\
\midrule
\cite{hyun2008seluge} &2008 &\begin{tabular}{@{}l@{}}Merkle hash tree signature verification \\of each transmitted block\end{tabular} &\begin{tabular}{@{}l@{}}- Pub Keys (IoT Devices) \\ - Secure Storage Technology for Private Key (FM)\end{tabular} \\
\midrule
\cite{samuel2010survivable} &2010 &Signature verification with multiple roles &\begin{tabular}{@{}l@{}}- Pub Keys (IoT Devices) \\ - Secure Storage Technology for Private Key (FM)\end{tabular} \\
\midrule
\cite{6289278} &2012 &\begin{tabular}{@{}l@{}}Elliptic Curve Cryptography (ECC) signature \\verification\end{tabular} &\begin{tabular}{@{}l@{}}- Pub Keys (IoT Devices) \\ - Secure Storage Technology for Private Key (FM)\end{tabular} \\
\midrule
\cite{salas2013secure} &2013 &Biometric-aided ECC cryptosystem &\begin{tabular}{@{}l@{}}- Dedicated Hardware (IoT devices) \\ - Secure Storage Technology (IoT devices)\end{tabular} \\
\midrule
\cite{doroodgar2014seluge++} &2014 &\begin{tabular}{@{}l@{}}Merkle hash tree signature verification \\of each transmitted block\end{tabular} &\begin{tabular}{@{}l@{}}- Pub Keys (IoT Devices) \\ - Secure Storage Technology for Private Key (FM)\end{tabular} \\
\midrule
\cite{7581459} &2016 &No integrity verification & \begin{tabular}{@{}c@{}} --- \end{tabular} \\
\midrule
\cite{karthik2016uptane} &2016 &Signature verification with multiple roles &\begin{tabular}{@{}l@{}}- Pub Keys (IoT Devices) \\ - Secure Storage Technology for Private Key (FM)\end{tabular} \\
\midrule
\cite{doddapaneni2017secure} &2017 &Signature Verification of FOTA Objects &\begin{tabular}{@{}l@{}}- Pub Keys (IoT Devices) \\ - Secure Storage Technology for Private Key (FM)\end{tabular} \\
\midrule
\cite{nikitin2017chainiac} &2017 &Collective signature verification &\begin{tabular}{@{}l@{}}- Pub Keys (IoT Devices) \\ - Secure Storage Technology for Private Key (FM) \\ - Secure connection with the Blockchain (IoT device)\end{tabular} \\
\midrule
\cite{lee2017blockchain} &2017 &\begin{tabular}{@{}l@{}}Signature verification of the metadata file stored \\in the Blockchain\end{tabular} &\begin{tabular}{@{}l@{}}- Storage space for the Blockchain (IoT devices) \\ - Secure connection with the Blockchain (IoT device) \\ - Pub Keys (nodes) \\ - Secure Storage Technology for Private Key (nodes) \end{tabular} \\
\midrule
\cite{teng2017firmware} &2017 &\begin{tabular}{@{}l@{}}Signature verification stored in Trusted \\Platform Modules\end{tabular} &\begin{tabular}{@{}l@{}}- Pub Keys (IoT Devices) \\ - Secure Storage Technology for Private Key (FM)\end{tabular} \\
\midrule
\cite{doddapaneni2017secure} &2017 &Signature and Encryption of Secure Object Format &\begin{tabular}{@{}l@{}}- Pub Keys (IoT Devices) \\ - Secure Storage Technology for Private Key (FM)\end{tabular} \\
\midrule
\cite{8016282} &2017 &Encrpytion using shared keys of the update payload &\begin{tabular}{@{}l@{}}- Physically Unclonable Functions (IoT Devices)\end{tabular} \\
\midrule
\cite{asokan2018assured} &2018 &RSA signature verification of the firmware metadata &- Hardware Security Module (all) \\
\midrule
\cite{yohan2018over} &2018 &FMart Contract signature verification of the firmware &\begin{tabular}{@{}l@{}}- Secure connection with the Blockchain (IoT device) \\ - Pub Keys (nodes) \\ - Secure Storage Technology for Private Key (nodes) \end{tabular} \\
\midrule
\cite{8719338} &2018 &Signature and Encryption using AES and RSA keys &\begin{tabular}{@{}l@{}}- Pub Keys (IoT Devices) \\- Secure Storage Technology for Private Key (FM) \\ - Secure Storage Technology for Shared Key (all)\end{tabular} \\
\midrule
\cite{zandberg2019secure} &2019 & \begin{tabular}{@{}l@{}}Elliptic Curve Cryptography (ECC) signature \\verification of manifest metadata\end{tabular} &\begin{tabular}{@{}l@{}}- Pub Keys (IoT Devices) \\ - Secure Storage Technology for Private Key (FM)\end{tabular} \\
\midrule
\cite{hu2019autonomous} &2019 &FMart Contract signature verification of the firmware &\begin{tabular}{@{}l@{}}- Hardware Security Module (all nodes) \\ - Secure connection with the Blockchain (IoT device)\end{tabular} \\
\midrule
\cite{8949023} &2019 &\begin{tabular}{@{}l@{}}Elliptic Curve Cryptography (ECC) signature \\verification of software update\end{tabular} &\begin{tabular}{@{}l@{}}- Pub Keys (IoT Devices) \\ - Secure Storage Technology for Private Key (FM)\end{tabular} \\
\midrule
\cite{mbakoyiannis2019secure} &2019 &\begin{tabular}{@{}l@{}}Multi-trust signature verification of manifest \\metadata\end{tabular} &\begin{tabular}{@{}l@{}}- Pub Keys (IoT Devices) \\ - Secure Storage Technology for Private Key (FM)\end{tabular} \\
\midrule
\cite{langiu2019upkit} &2019 &Double signature on the update image &\begin{tabular}{@{}l@{}}- Pub Keys (IoT Devices) \\ - Secure Storage Technology for Private Key (FM)\end{tabular} \\
\midrule
\cite{8959992} &2019 &Hardware-based signature verification &\begin{tabular}{@{}l@{}}- Pub Keys (IoT Devices) \\ - Secure Storage Technology (IoT devices) \\ - Dedicated Hardware (IoT devices)\end{tabular} \\
\midrule
\cite{9059463} &2019 &Hardware-based signature verification &\begin{tabular}{@{}l@{}}- Secure Storage Technology (IoT devices) \\ - Dedicated Hardware (IoT devices)\end{tabular} \\
\midrule
\cite{8728389} &2019 &Verification of Hash Chain &\begin{tabular}{@{}l@{}}- Storage space for the Blockchain (IoT devices) \\ - Secure connection with the Blockchain (IoT device) \\ - Pub Keys (nodes) \\ - Secure Storage Technology for Private Key (nodes)\end{tabular} \\
\midrule
\cite{8673027} &2019 &\begin{tabular}{@{}l@{}}Integrity verification executed in the Blockchain \\Server\end{tabular} &\begin{tabular}{@{}l@{}}- Tamper-proof Blockchain server \\ - Secure connection with the Blockchain Server (IoT device)\end{tabular} \\
\midrule
\cite{8939910} &2019 & \begin{tabular}{@{}l@{}}Peer-to-peer update with Smart Contract signature \\verification of the firmware\end{tabular} &\begin{tabular}{@{}l@{}}- Secure connection with the Blockchain (IoT device) \\ - Pub Keys (nodes) \\ - Secure Storage Technology for Private Key (nodes)\end{tabular} \\
\midrule
\cite{yohan2020fotb} &2020 &Peer-to-peer verification process through consensus &\begin{tabular}{@{}l@{}}- Pub Keys (IoT Devices) \\ - Secure Storage Technology for Private Key (FM)\end{tabular} \\
\midrule
\cite{anastasiou2020iot} &2020 &Smart Contract signature verification of the firmware &\begin{tabular}{@{}l@{}}- Hardware Security Module (all nodes) \\ - Secure connection with the Blockchain (IoT device)\end{tabular} \\
\midrule
\cite{sahlmann2020mup} &2020 &Digest signature verification &\begin{tabular}{@{}l@{}}- Pub Keys (IoT Devices) \\ - Secure Storage Technology for Private Key (FM)\end{tabular} \\
\midrule
\cite{falas2021modular} &2021 &Hardware-based signature verification &\begin{tabular}{@{}l@{}}- Pub Keys (IoT Devices) \\ - Secure Storage Technology for Private Key (FM)\end{tabular} \\
\midrule
\cite{tsaur2022highly} &2022 &Smart Contract signature verification of the firmware &\begin{tabular}{@{}l@{}}- Storage space for the Blockchain (IoT devices) \\ - Secure connection with the Blockchain (IoT device) \\ - Pub Keys (nodes) \\ - Secure Storage Technology for Private Key (nodes)\end{tabular} \\
\midrule
\cite{ghosal2022secure} &2022 &Digest signature verification &\begin{tabular}{@{}l@{}}- Pub Keys (IoT Devices) \\ - Secure Storage Technology for Private Key (FM)\end{tabular} \\
\midrule
\cite{de2022over} &2022 &Digest signature verification &\begin{tabular}{@{}l@{}}- Pub Keys (IoT Devices) \\ - Secure Storage Technology for Private Key (FM)\end{tabular} \\
\bottomrule

\caption{Analysis of the integrity protection mechanisms adopted by state-of-the-art firmware update solutions.}
\label{tab:sota}

\end{longtable}
\end{scriptsize}

\section{\methodname{}}
\label{sec:methodology}

The analysis of the threats to the firmware integrity, the definition of the attacker model, and the evaluation of the main solutions to ensure the integrity of the IoT firmware during the update process allowed us to determine that:

\begin{enumerate}
    \item state-of-the-art solutions are vulnerable to IMG.MODIFICATION, i.e., an attacker or a malicious FM can repack the content of the IoT firmware before the signing process in the generation phase;
    \item All solutions based on the signature of the firmware bundle or the associated metadata are vulnerable to the KEY.EXPOSURE and META.TOCTOU threats thereby vanishing the adopted enforcement mechanisms; the only way to cope with such threats is to adopt secure storage technologies or hardware security modules that may not be compatible with low-end IoT devices;
\end{enumerate}

In this section, we introduce the basics of \methodname{} (\underline{P}ervasive \underline{A}nti-\underline{R}epackaging for \underline{IoT}), the first solution of internal anti-repackaging for IoT firmware that can extend existing software update solutions to ensure resilience against repackaging attacks in the whole production and delivery process. \methodname{} does not require signing keys, secure storage technologies, or hardware security modules to ensure full compatibility with low-end IoT devices.

\methodname{} protects an IoT firmware by injecting self-protecting code directly inside the firmware code.
The methodology exploits the use of Cryptographically obfuscated Logic Bombs \cite{10.1145/3168820} (CLB) to hide anti-tampering (AT) checks in the firmware executable. Such CLBs will be triggered (i.e., \textit{explode}) in case of any tampering is detected.
Moreover, to defuse an AT check inside a CLB, the attacker would need to execute the CLB, retrieve the value of the decryption key to decrypt its content, and then bypass or remove the AT checks injected in the body of the qualified condition.

This section details the \methodname{} protection scheme and its runtime behavior.

\subsection{Protection Scheme and Runtime Behaviour}

The \methodname{} protection scheme is based on the dissemination in the IoT firmware of CLBs that hide a set of AT checks. 
To minimize the complexity, each CLB embeds a single AT check that performs a signature verification of a portion of the IoT firmware executable. 
Still, it is worth noticing that \methodname{} supports the definition of other AT checks as well as different CLB schemes.

\begin{figure*}[!ht]
  \includegraphics[width=\textwidth]{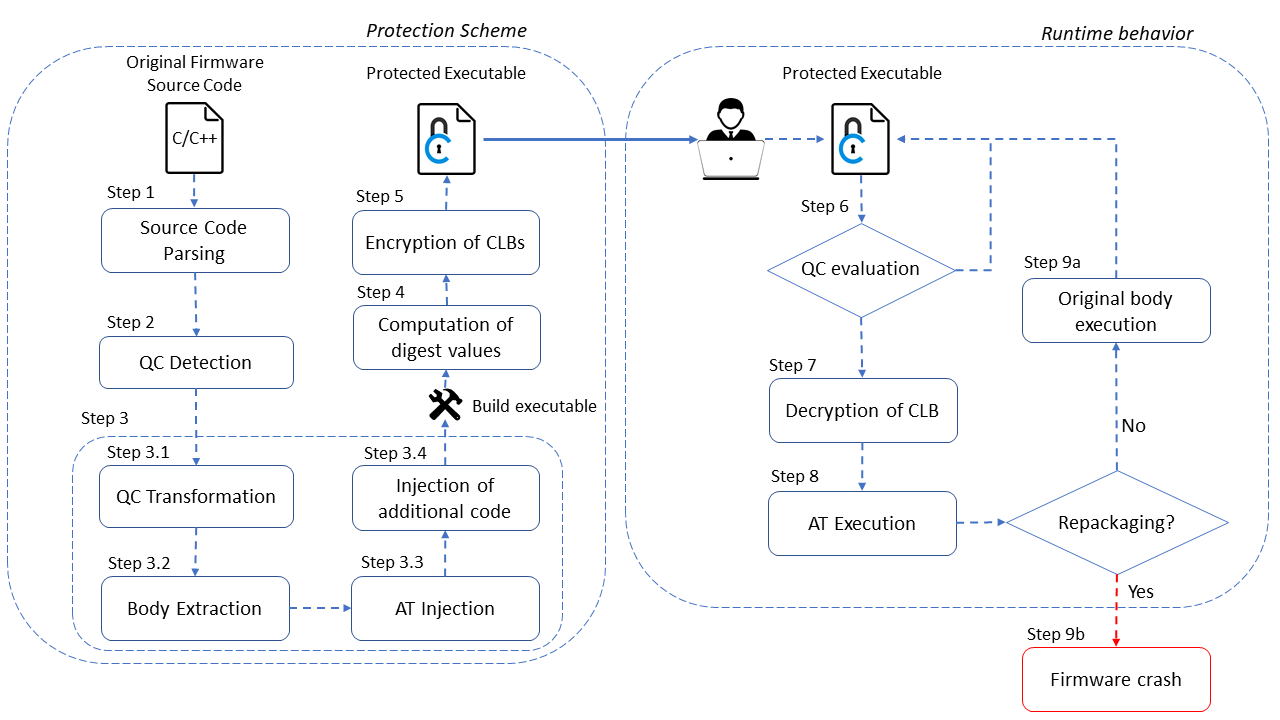}
  \caption{High-level overview of \methodname{} protection process and runtime behavior of a protected firmware.}
  \label{fig:IoT_antirepackaging_process}
\end{figure*}

Figure \ref{fig:IoT_antirepackaging_process} shows a high-level overview of the proposed technique's protection process and the runtime behavior.
\methodname{} starts the protection process from the source code of the IoT firmware. 
In detail, it parses the source code (Step 1) to identify the set of suitably qualified conditions (Step 2).
In Step 3, these conditions are transformed according to the CLB schema.
During this process, the equality check of the QC is transformed into the cryptographically obfuscated form (Step 3.1). Then, \methodname{} extracts the body of the qualified condition (Step 3.2), injects the AT code (Step 3.3), and adds the logic to decrypt and execute the new body once in the encrypted form (Step 3.4).
The choice of injecting the CLBs directly into the source code allows for minimizing invasive procedures (e.g., binary rewriting) that may break the compatibility with existing firmware update frameworks or disrupt the firmware image.

Depending on the implemented AT controls, the scheme could compute some digest values (e.g., the hash code for some executable code), storing their values inside the corresponding bombs. 
To do so, \methodname{} triggers the firmware compilation process to complete the protection.
This process depends on the firmware being protected: \methodname{} simply invokes its  original build system.
In Steps 4-5, the computed values are the expected results of each AT, which executes the integrity check by comparing a runtime computed value with the digest stored in the bomb.

Finally, \methodname{} encrypts each CLB with their constant values (\texttt{const}) to obtain the protected firmware.

At runtime, the code of the CLBs is executed iff the value of the variable in the QC (plus a salt) is equal to the constant value (i.e., if its hash matches \texttt{Hconst} - Step 6). If this is the case, the body of the CLB is decrypted using the \texttt{const} value as the decryption key (Step 7), and the corresponding code is executed. This behavior triggers the AT check (Step 8) that computes the digest of a portion of the firmware executable and compares it with the stored one. If these values match, the execution can proceed normally (Step 9a). Otherwise, the AT reports a tampering attempt and executes an action, like sending an alert to the Firmware Manufacturer or triggering a Security Exception and aborting the execution (i.e., the case of Step 9b). 
\section{\toolname{}}
\label{sec:implementation}
To demonstrate the applicability and the feasibility of \methodname{}, we developed \toolname{} (i.e., \underline{\methodname} for \underline{I}ntegrated \underline{C}-based firmware) to support the protection of IoT firmware designed in C/C++ programming language. The tool is publicly available on GitHub at \cite{patriot}.

\toolname{} consists of two main modules:

\begin{itemize}
    \item \moduleone{}. This module works directly on the firmware source code and is responsible for parsing of the source code, detect the QCs, and build of CLBs (Steps 1-3 of the protection process). 
    \item \moduletwo{}. This module processes the compiled IoT firmware and is responsible for computing the signature-verification digests of the AT checks and encrypting the CLBs (Steps 4-5 of Figure \ref{fig:IoT_antirepackaging_process}).
\end{itemize}

\subsection{\moduleone{}}

\moduleone{} is built using the Python language and leverages the Clang library\footnote{https://github.com/llvm-mirror/clang/tree/master/bindings/python} to pre-process the C/C++ source code.

During this phase, \moduleone{} scans the source code to obtain the list of qualified conditions that can host a logic bomb. 
In the current implementation, \moduleone{} supports if-then-else statements with an equality condition of the form of \texttt{X == const}. 
If we consider the source code of Listing \ref{lst:c_original_code} as an example, \moduleone{} would detect one QC at row \ref{line:qc}.
\begin{lstlisting}[caption={Example of a C source code.}, label={lst:c_original_code}]
int funA(int val);

void funB(int val) {
    int a = 0;
    [...]
    /* QC */
    if (a == CONST) { |\label{line:qc}|
        /* or_code */
        int res = funA(val);
        printf("The result is %d", res);
    }
    [...]
}
\end{lstlisting}

After the QC detection phase, \moduleone{} converts each QC in the corresponding CLB.
To do so, the tool computes the hash of the constant value of the QC, generates a 4-bytes random salt, and modifies the \texttt{if} condition to match the form \texttt{if(H(X,salt) == Hconst)}. 
Then, it creates a new function ($ext\_fun$) that encapsulates the QC original body and accepts -- as input parameters -- all the variables used inside the body.
Moreover, during this phase, the module adds an AT control in $ext\_fun$ (Step 3.3 of Figure \ref{fig:IoT_antirepackaging_process}). 
In the current implementation, the AT control evaluates the hash signature of a portion of the compiled firmware and raises a security exception in case of a signature mismatch.

Since \moduleone{} works directly on the source code, it injects three placeholder values into the source code that will be updated by \moduletwo{} before the encryption phase. In detail, the module adds three variables, i.e., \texttt{offset}, and \texttt{count} to identify the part of the executable to evaluate in the AT control, and \texttt{control\_value} for the expected result of the verification.

Finally, \moduleone{} injects the functions to decrypt and execute $ext\_fun$ in the body of the CLB.

Listing \ref{lst:c_protected_code} reports the processing result of \moduleone{} on the example code of Listing \ref{lst:c_original_code}. 
Starting from the QC located in $funB$, \moduleone{} creates a new function ($ext\_funB$) that contains the original body of the QC (rows \ref{line:ext-start}-\ref{line:ext-end}). Then, it injects an anti-tampering control (row \ref{line:at}) that verifies a portion of the executable file (identified by \texttt{offset} and \texttt{count} - rows \ref{line:off} and \ref{line:count}) against the expected hash value (i.e., the variable \texttt{control\_value} - row \ref{line:control}).
Finally, \moduleone{} replaces the original body of the QC with the code to decrypt and execute the original function (rows \ref{line:dec} and \ref{line:exe}).

\definecolor{amber}{rgb}{1.0, 0.75, 0.0}
\begin{lstlisting}[caption={Output code produced by \moduleone{}.}, label={lst:c_protected_code}]
int funA(int value);

/* ptr to start of the firmware */
uint8_t *ptr;

void at_check(off_t offset, size_t count, int control_value, uint8_t *ptr) {
    int i; 
    uint8_t *buf = malloc(sizeof(uint8_t) * count);
    /* Read the bytes to control */
    if (i = 0; i < count; i++) {
        buf[i] = ptr[offset+i];
    }
    /* Integrity check */
    if (hash(buf) != control_value) {
        puts("*Aborting: Security Exception (Repackaging detected)");
        free(buf);
        exit(123);
    }
    free(buf);
    return;
}

void ext_funB(int* val) { |\label{line:ext-start}|
    /* Placeholders */ 
    off_t offset = |\textcolor{red}{0x0ff53701}| ; |\label{line:off}|
    size_t count = |\textcolor{orange}{0xb17e5010}|; |\label{line:count}|
    int control_value = |\textcolor{amber}{0x4559ffff}|; |\label{line:control}|
    /* AT check */
    at_check(offset, count, control_value, ptr); |\label{line:at}|
    /* or_code of funB */
    int res = funA(val);
    printf("The result is %d\n", res);
} |\label{line:ext-end}|

void funB(int val) {
    int a = 0;
    [...]
    /* CLB */
    if (hash(a,salt) == Hconst) {
        decrypt(&ext_funB, &a); |\label{line:dec}|
        ext_funB(&val); |\label{line:exe}|
    }
    [...]
}
\end{lstlisting}

\subsection{\moduletwo{}}

\moduletwo{} is a Java command-line tool that processes the compiled IoT firmware to i) update the control values of the AT checks and ii) encrypt the content of the CLBs.
The module receives from \moduleone{} the list of CLBs, the functions that need encryption (i.e., the list of $ext\_fun$ methods), and their corresponding encryption keys (i.e., the $const$ values).

For each CLB, the tool exploits \texttt{nm}\footnote{\url{https://linux.die.net/man/1/nm}} to locate in the firmware executable the corresponding $ext\_fun$ and the position of the embedded control values (i.e., \texttt{offset}, \texttt{count}, and \texttt{control\_value}).
Also, the module identifies the portion of code that the AT checks will evaluate. 
The current version of \moduletwo{} selects all the compiled code (the content of the \texttt{.text} elf section).
From the selected code, the module computes: i) the starting position (\texttt{offset}), ii) the number of bytes (\texttt{count}), and iii) the hash of the selection (\texttt{control\_value}). Then, \moduletwo{} replaces the placeholder values with the obtained results (Step 4 of Figure \ref{fig:IoT_antirepackaging_process}).
Finally, \moduletwo{} encrypts the bytes of the $ext\_fun$ using the $const$ value as the encryption key (step 5).

\begin{figure}[h!]
    \centering
    \subcaptionbox{Identification of the control values of $ext\_funB$. \label{fig:elf-loc}}{
        \includegraphics[width=0.64\linewidth]{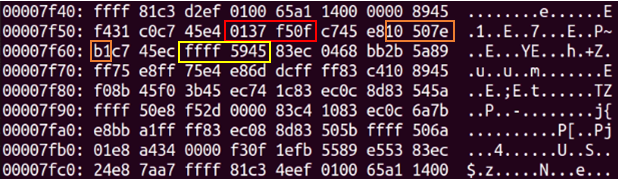}
    }%
    \hfill

    \subcaptionbox{Computation and updating of the control values of $ext\_funB$. \label{fig:elf-upd}}{
        \includegraphics[width=0.64\linewidth]{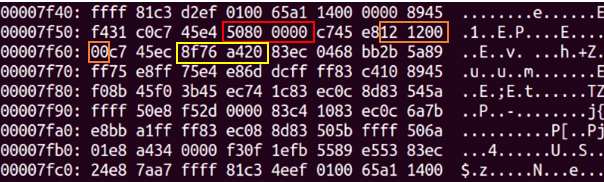}
    }%
    \hfill

    \subcaptionbox{Encryption of $ext\_funB$. \label{fig:elf-enc}}{
        \includegraphics[width=0.64\linewidth]{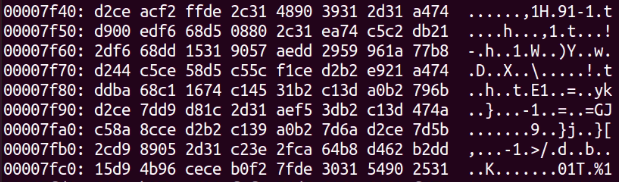}
    }%
    \caption{\moduletwo{} protection process on the part of the executable file containing $ext\_funB$ of Listing \ref{lst:c_protected_code}.}
    \label{fig:elf_protection_example}
\end{figure}

Figure \ref{fig:elf_protection_example} shows the protection applied by \moduletwo{} on the part of the executable file containing $ext\_funB$ of Listing \ref{lst:c_protected_code}. In detail, the tool locates the control values (Figure \ref{fig:elf-loc}), computes and updates their values (Figure \ref{fig:elf-upd}) and, then, encrypts the entire function (Figure \ref{fig:elf-enc}).

\section{Experimental Evaluation}
\label{sec:experiments}

We empirically assessed the applicability of \methodname{} by applying the \toolname{} protection on 50 real-world samples for resource-constrained IoT devices in a RIOT-based ecosystem with the SUIT update framework.

RIOT is an open-source OS designed for resource-constrained IoT devices that have gained the scientific community's attention in the last few years~\cite{baccelli2018riot}.
RIOT allows for standard C and C++ application programming, provides multi-threading and real-time capabilities, and only requires a minimum of 1.5 KB of RAM. RIOT consists of a microkernel architecture with core functionalities and pluggable modules that support multiple network stacks, libraries, and utility \cite{modules}.

In detail, the experimental campaign applied the tool on each firmware to evaluate the distribution of the protection controls and the introduced size overhead.
Each firmware is built using RIOT OS version 2021.05 and a different RIOT app (as detailed in Section \ref{sec:dataset}) for two different boards, i.e., \textit{native} \cite{native} and \textit{iotlab-m3} \cite{m3}.
Hereafter, we refer to each firmware with the name of the contained application.

The building and protection phase was performed on a virtual machine running Ubuntu 20.04 with four processors and 16GB RAM.

Then, we evaluated the reliability of the protection scheme at runtime. First, we executed the protected firmware images to check the proper functioning. Then, we tested the solution against actual repackaging attacks by attempting to repackage each protected firmware and testing the tampered file. The runtime evaluations were executed on a real \textit{iotlab-m3 board} hosted by the FIT IoT-LAB testbed~\cite{adjih2015fit}. The iotlab-m3 board has an STM32 MCU, 32-bit Cortex M3 CPU, 64 KB of RAM, and 256 KB of ROM.

\subsection{Dataset}
\label{sec:dataset}
The preparation of the dataset consisted of the following steps.
First, we scraped GitHub~\cite{github} looking for recent repositories (i.e., since January 2019) that contain the words ``RIOT OS'' and ``IoT'' and at least one of the following keywords: \{``app'', ``application'', ``firmware'', ``code''\} (e.g., query \texttt{RIOT OS IoT firmware site:github.com after:2019-1-1}).
The analysis resulted in the identification of 150 unique public GitHub repositories matching our criteria.
From these, we manually examined the collected repositories to retain only the ones that contain IoT apps compatible with recent RIOT OS versions (i.e., at least $>=$ 2021.05) and support the IoT boards included in the experimental evaluation (i.e., \textit{native} and \textit{iotlab-m3}).
After the review process, we obtained a set of 50 different apps distributed across 16 GitHub repositories ($\sim$ 10\% -- $16 / 150$). Table~\ref{tbl:riotapps} in ~\ref{app:dataset} reports the name of each app, the link to the GitHub project, and the subfolder that contains the source code.
Finally, we built each firmware using RIOT OS version 2021.05 and a different RIOT app for the \textit{native} and \textit{iotlab-m3} boards. 

It is important to notice that the low availability of open-source and working apps for the RIOT OS directly influenced the magnitude of the dataset. Still, we obtained a representative collection of samples as the apps come from heterogeneous scenarios and use different RIOT modules.
Notables examples are the \texttt{museum} app, which is part of the \textit{ARte} (\textit{Augmented Reality to educate}) project, and the \texttt{election\_master} app.
The former leverages the MQTT protocol~\cite{mqtt} relying on the \texttt{emcute} RIOT module to improve the interaction between visitors and artworks with the COVID-19 restrictions. The latter implements a custom leader election algorithm on RIOT OS nodes connected through the network and includes the modules for the routing protocol (i.e., \texttt{gnrc\_rpl}, and \texttt{auto\_init\_gnrc\_rpl}).
We reported in Table \ref{tbl:app2module} in \ref{app:dataset} a detail of the RIOT modules used in each app of the dataset.
Finally, It is worth empathizing that large-scale analysis is out of the scope of this work: our evaluation aims to demonstrate the effectiveness and the enforceability of \toolname{} regardless of the features (i.e., included RIOT OS modules) of the RIOT apps.

\subsection{Protection Evaluation} 

\toolname{} was able to apply the protections over the entire dataset in nearly 53 minutes (i.e., 3208 seconds) for iotlab-m3 board and 46 minutes (2818 seconds) for the native board. 
The protection of a single firmware took, on average, 64.2 seconds. 

\toolname{}  worked successfully in 90\% (45/50) of the cases, i.e., it generated valid protected firmware. The remaining 10\% (i.e., 5 firmware images based on \texttt{decho}, \texttt{dsock}, \texttt{gnrc\_brouter}, \texttt{nanocoap}, and \texttt{ndn} apps) failed due to errors in the building phase. We manually investigated such problems to discover that the build process failed due to the presence of at least an \textit{ext\_fun} with i) unsupported instructions (e.g., goto statements to undefined portions of code), or ii) undefined variables. 
These problems are mainly attributable to the parsing of the source code (i.e., Step 1 of Figure \ref{fig:IoT_antirepackaging_process}) that leads to incorrect identification of the body of the QCs by the Clang python extension.

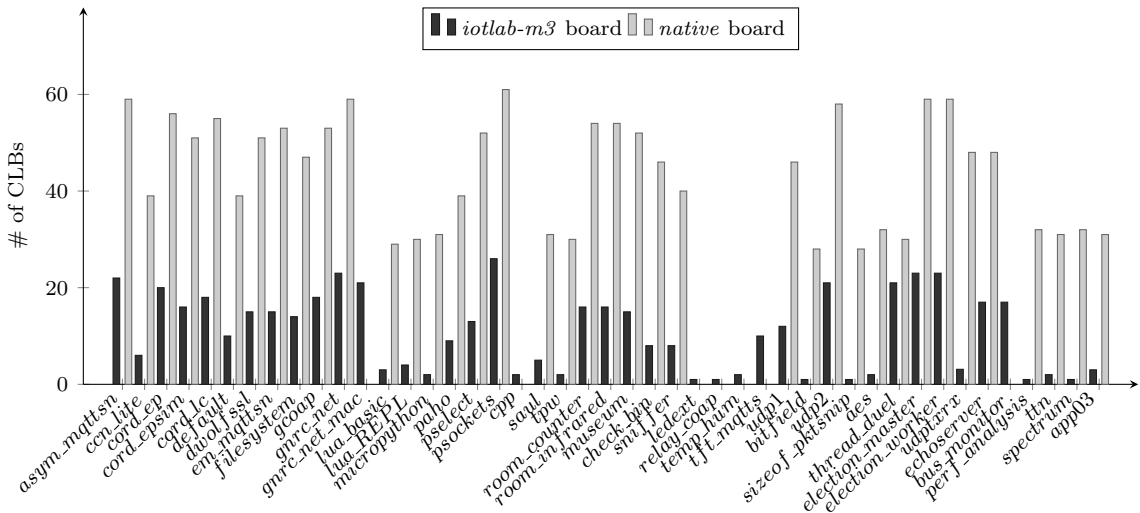
\begin{figure}[h]
    \centering
    \begin{tikzpicture}[
      every axis/.style={ 
        ybar, 
        ymin=0, ymax=78,
        axis lines=left,
        enlarge y limits=0,
        enlarge x limits=0.04,
        legend style={at={(0.4,-0.35)},
            anchor=north,legend columns=-1},
        ylabel={\# of CLBs},
        x tick label style={rotate=45,anchor=east},
        symbolic x coords={
        ${asym\_mqttsn}$,
        ${ccn\_lite}$,
        ${cord\_ep}$,
        ${cord\_epsim}$,
        ${cord\_lc}$,
        ${default}$,
        ${dwolfssl}$,
        ${em\_mqttsn}$,
        ${filesystem}$,
        ${gcoap}$,
        ${gnrc\_net}$,
        ${gnrc\_net\_mac}$,
        ${lua\_basic}$,
        ${lua\_REPL}$,
        ${micropython}$,
        ${paho}$,
        ${pselect}$,
        ${psockets}$,
        ${cpp}$,
        ${saul}$,
        ${tpw}$,
        ${room\_counter}$,
        ${room\_infrared}$,
        ${museum}$,
        ${check\_bin}$,
        ${sniffer}$,
        ${ledext}$,
        ${relay\_coap}$,
        ${temp\_hum}$,
        ${tft\_mqtts}$,
        ${udp1}$,
        ${bitfield}$,
        ${udp2}$,
        ${sizeof\_pktsnip}$,
        ${aes}$,
        ${thread\_duel}$,
        ${election\_master}$,
        ${election\_worker}$,
        ${udptxrx}$,
        ${echoserver}$,
        ${bus\_monitor}$,
        ${perf\_analysis}$,
        ${ttn}$,
        ${spectrum}$,
        ${app03}$
        },
        xtick={
        ${asym\_mqttsn}$,
        ${ccn\_lite}$,
        ${cord\_ep}$,
        ${cord\_epsim}$,
        ${cord\_lc}$,
        ${default}$,
        ${dwolfssl}$,
        ${em\_mqttsn}$,
        ${filesystem}$,
        ${gcoap}$,
        ${gnrc\_net}$,
        ${gnrc\_net\_mac}$,
        ${lua\_basic}$,
        ${lua\_REPL}$,
        ${micropython}$,
        ${paho}$,
        ${pselect}$,
        ${psockets}$,
        ${cpp}$,
        ${saul}$,
        ${tpw}$,
        ${room\_counter}$,
        ${room\_infrared}$,
        ${museum}$,
        ${check\_bin}$,
        ${sniffer}$,
        ${ledext}$,
        ${relay\_coap}$,
        ${temp\_hum}$,
        ${tft\_mqtts}$,
        ${udp1}$,
        ${bitfield}$,
        ${udp2}$,
        ${sizeof\_pktsnip}$,
        ${aes}$,
        ${thread\_duel}$,
        ${election\_master}$,
        ${election\_worker}$,
        ${udptxrx}$,
        ${echoserver}$,
        ${bus\_monitor}$,
        ${perf\_analysis}$,
        ${ttn}$,
        ${spectrum}$,
        ${app03}$
        },
        width=0.94\textwidth, height=0.40\textwidth,
        legend style={at={(0.5,1)},
                anchor=north,legend columns=-1}
      },
    ]
    
    \tikzstyle{every node}=[font=\footnotesize]
    \begin{axis}
        \addplot+[gray!20!black,fill=black!80!white, bar width=2.5pt, y=0] coordinates
        {(${asym\_mqttsn}$,22)
        (${ccn\_lite}$,6)
        (${cord\_ep}$,20)
        (${cord\_epsim}$,16)
        (${cord\_lc}$,18)
        (${default}$,10)
        (${dwolfssl}$,15)
        (${em\_mqttsn}$,15)
        (${filesystem}$,14)
        (${gcoap}$,18)
        (${gnrc\_net}$,23)
        (${gnrc\_net\_mac}$,21) 
        (${lua\_basic}$,3)
        (${lua\_REPL}$,4)
        (${micropython}$,2)
        (${paho}$,9)
        (${pselect}$,13)
        (${psockets}$,26)
        (${cpp}$,2)
        (${saul}$,5)
        (${tpw}$,2)
        (${room\_counter}$,16)
        (${room\_infrared}$,16)
        (${museum}$,15)
        (${check\_bin}$,8) 
        (${sniffer}$,8)
        (${ledext}$,1)
        (${relay\_coap}$,1)
        (${temp\_hum}$,2) 
        (${tft\_mqtts}$,10)
        (${udp1}$,12)
        (${bitfield}$,1)
        (${udp2}$,21)
        (${sizeof\_pktsnip}$,1)
        (${aes}$,2)
        (${thread\_duel}$,21)
        (${election\_master}$,23)
        (${election\_worker}$,23)
        (${udptxrx}$,3.11)
        (${echoserver}$,17)
        (${bus\_monitor}$,17)
        (${perf\_analysis}$,1)
        (${ttn}$,2)
        (${spectrum}$,1)
        (${app03}$,3)};
    \addplot+[gray!80!black,fill=black!20!white, bar width=2.5pt, y=1] coordinates
        {(${asym\_mqttsn}$,59)
        (${ccn\_lite}$,39)
        (${cord\_ep}$,56)
        (${cord\_epsim}$,51)
        (${cord\_lc}$,55)
        (${default}$,39)
        (${dwolfssl}$,51)
        (${em\_mqttsn}$,53)
        (${filesystem}$,47)
        (${gcoap}$,53)
        (${gnrc\_net}$,59)
        (${lua\_basic}$,29)
        (${lua\_REPL}$,30)
        (${micropython}$,31)
        (${paho}$,39)
        (${pselect}$,52)
        (${psockets}$,61)
        (${saul}$,31)
        (${tpw}$,30)
        (${room\_counter}$,54)
        (${room\_infrared}$,54)
        (${museum}$,52)
        (${check\_bin}$,46) 
        (${sniffer}$,40)
        (${udp1}$,46)
        (${bitfield}$,28)
        (${udp2}$,58)
        (${sizeof\_pktsnip}$,28)
        (${aes}$,32)
        (${thread\_duel}$,30)
        (${election\_master}$,59)
        (${election\_worker}$,59)
        (${udptxrx}$,48)
        (${echoserver}$,48)
        (${perf\_analysis}$,32)
        (${ttn}$,31)
        (${spectrum}$,32)
        (${app03}$,31)};

       \addlegendentry{\textit{iotlab-m3} board}
       \addlegendentry{\textit{native} board}
    \end{axis}
    \end{tikzpicture}
    
    \caption{Number of  CLBs injected in the RIOT firmware samples of the dataset.} 
    \label{fig:distribution-lbs}
\end{figure}

Figure \ref{fig:distribution-lbs} shows the number of logic bombs distributed in each protected RIOT firmware for the two boards.
\toolname{} injected, on average, 10.9 CLBs (st. dev. 8.1) on firmware images for iotlab-m3 and 43 CLBs (st. dev. 11.9) on the ones compatible with the native board. 
The significant difference in injected bombs reflects the peculiarities of the two target devices regarding the codebase, compatible APIs, and external modules included by default. In particular, the native board encloses additional code and, consequently, more QCs to apply the protection scheme concerning the Cortex-based counterpart. 
Finally, it is worth noting that 12 executables for iotlab-m3 boards have few CLBs (i.e., less than 3) since they include a basic app (e.g.,  similar to \textit{hello world}), thereby exacerbating such a gap.

\begin{figure}[h]
    \centering
    \begin{tikzpicture}
    \tikzstyle{every node}=[font=\scriptsize]
    \begin{axis}[
        ylabel={Size overheads (\%)},
        xlabel={RIOT firmware},
        ymin=-1, ymax=20,
        x tick label style={rotate=45,anchor=east},
        symbolic x coords={
        ${asym\_mqttsn}$,
        ${ccn\_lite}$,
        ${cord\_ep}$,
        ${cord\_epsim}$,
        ${cord\_lc}$,
        ${default}$,
        ${dwolfssl}$,
        ${em\_mqttsn}$,
        ${filesystem}$,
        ${gcoap}$,
        ${gnrc\_net}$,
        ${gnrc\_net\_mac}$,
        ${lua\_basic}$,
        ${lua\_REPL}$,
        ${micropython}$,
        ${paho}$,
        ${pselect}$,
        ${psockets}$,
        ${cpp}$,
        ${saul}$,
        ${tpw}$,
        ${room\_counter}$,
        ${room\_infrared}$,
        ${museum}$,
        ${check\_bin}$,
        ${sniffer}$,
        ${ledext}$,
        ${relay\_coap}$,
        ${temp\_hum}$,
        ${tft\_mqtts}$,
        ${udp1}$,
        ${bitfield}$,
        ${udp2}$,
        ${sizeof\_pktsnip}$,
        ${aes}$,
        ${thread\_duel}$,
        ${election\_master}$,
        ${election\_worker}$,
        ${udptxrx}$,
        ${echoserver}$,
        ${bus\_monitor}$,
        ${perf\_analysis}$,
        ${ttn}$,
        ${spectrum}$,
        ${app03}$
        },
        xtick={
        ${asym\_mqttsn}$,
        ${ccn\_lite}$,
        ${cord\_ep}$,
        ${cord\_epsim}$,
        ${cord\_lc}$,
        ${default}$,
        ${dwolfssl}$,
        ${em\_mqttsn}$,
        ${filesystem}$,
        ${gcoap}$,
        ${gnrc\_net}$,
        ${gnrc\_net\_mac}$,
        ${lua\_basic}$,
        ${lua\_REPL}$,
        ${micropython}$,
        ${paho}$,
        ${pselect}$,
        ${psockets}$,
        ${cpp}$,
        ${saul}$,
        ${tpw}$,
        ${room\_counter}$,
        ${room\_infrared}$,
        ${museum}$,
        ${check\_bin}$,
        ${sniffer}$,
        ${ledext}$,
        ${relay\_coap}$,
        ${temp\_hum}$,
        ${tft\_mqtts}$,
        ${udp1}$,
        ${bitfield}$,
        ${udp2}$,
        ${sizeof\_pktsnip}$,
        ${aes}$,
        ${thread\_duel}$,
        ${election\_master}$,
        ${election\_worker}$,
        ${udptxrx}$,
        ${echoserver}$,
        ${bus\_monitor}$,
        ${perf\_analysis}$,
        ${ttn}$,
        ${spectrum}$,
        ${app03}$
        },
        ymajorgrids=true,
        grid style=dashed,
        enlarge x limits=0.04,
        width=0.94\textwidth, height=0.22\textwidth,
        legend style={at={(0.5,1.4)},
                anchor=north,legend columns=-1}
    ]
    
    \addplot[
        color=blue,
        mark=square,
        ]
        coordinates {
        (${asym\_mqttsn}$,4.878860582190356)
        (${ccn\_lite}$,1.6854028898409823)
        (${cord\_ep}$,4.695674566685333)
        (${cord\_epsim}$,4.307642976829446)
        (${cord\_lc}$,3.9924736517309394)
        (${default}$,3.113301112496419)
        (${dwolfssl}$,2.7706892113482824)
        (${em\_mqttsn}$,4.401620946310786)
        (${filesystem}$,4.3210969923726115)
        (${gcoap}$,4.04241561500429)
        (${gnrc\_net}$,4.737407428345299)
        (${gnrc\_net\_mac}$,6.427407428345299) 
        (${lua\_basic}$,0.5581922729976143)
        (${lua\_REPL}$,0.5745497506046149)
        (${micropython}$,0.5518167665969547)
        (${paho}$,2.3729881878101664)
        (${pselect}$,4.408520790937354)
        (${psockets}$,4.37763465224178)
        (${cpp}$,1.8866384487121186)
        (${saul}$,2.5661882331482206)
        (${tpw}$,3.341061326015397)
        (${room\_counter}$,3.201250134022156)
        (${room\_infrared}$,3.208367247876507)
        (${museum}$,2.9729626211295086)
        (${check\_bin}$,3.8329626211295086) 
        (${sniffer}$,2.961995903750105)
        (${ledext}$,3.339700786070493)
        (${relay\_coap}$,3.3396324004178344)
        (${temp\_hum}$,4.4896324004178344) 
        (${tft\_mqtts}$,3.477195223226569)
        (${udp1}$,3.876879803163362)
        (${bitfield}$,5.723813910640258)
        (${udp2}$,4.755238417182302)
        (${sizeof\_pktsnip}$,5.978892506262227)
        (${aes}$,2.3680930619027833)
        (${thread\_duel}$,2.064876132853711)
        (${election\_master}$,3.927792016927619)
        (${election\_worker}$,3.8995968889362547)
        (${udptxrx}$,3.5144940230668533)
        (${echoserver}$,3.442346551757109)
        (${bus\_monitor}$,3.759075584738463)
        (${perf\_analysis}$,3.307380735495477)
        (${ttn}$,3.1748452414668837)
        (${spectrum}$,2.7738909296411416)
        (${app03}$,3.2537404288587437)
 };
    \addplot[
        color=red,
        mark=triangle,
        ]
        coordinates {
        (${asym\_mqttsn}$,11.9)
        (${ccn\_lite}$,9.2)
        (${cord\_ep}$,10.7)
        (${cord\_epsim}$,10.7)
        (${cord\_lc}$,10.6)
        (${default}$,11.9)
        (${dwolfssl}$,7.4)
        (${em\_mqttsn}$,11.4)
        (${filesystem}$,13.7)
        (${gcoap}$,10.9)
        (${gnrc\_net}$,11.3)
        (${lua\_basic}$,6.1)
        (${lua\_REPL}$,6.5)
        (${micropython}$,3.5)
        (${paho}$,6.1)
        (${pselect}$,11.2)
        (${psockets}$,11.2)
        (${saul}$,15.3)
        (${tpw}$,14.8)
        (${room\_counter}$,11.5)
        (${room\_infrared}$,11.4)
        (${museum}$,10.7)
        (${check\_bin}$,7.4) 
        (${sniffer}$,9.1)
        (${udp1}$,9.9)
        (${bitfield}$,15.0)
        (${udp2}$,11.0)
        (${sizeof\_pktsnip}$,15.7)
        (${aes}$,13.9)
        (${thread\_duel}$,14.5)
        (${election\_master}$,10.0)
        (${election\_worker}$,9.9)
        (${udptxrx}$,10.1)
        (${echoserver}$,10.1)
        (${perf\_analysis}$,13.6)
        (${ttn}$,13.7)
        (${spectrum}$,10.6)
        (${app03}$,14.3)
        };

        \addlegendentry{\textit{iotlab-m3} board}
       \addlegendentry{\textit{native} board}
    \end{axis}
    \end{tikzpicture}
    
    \caption{Distribution of the size overhead over the protected RIOT apps.}
    \label{fig:size-overhead}
\end{figure}
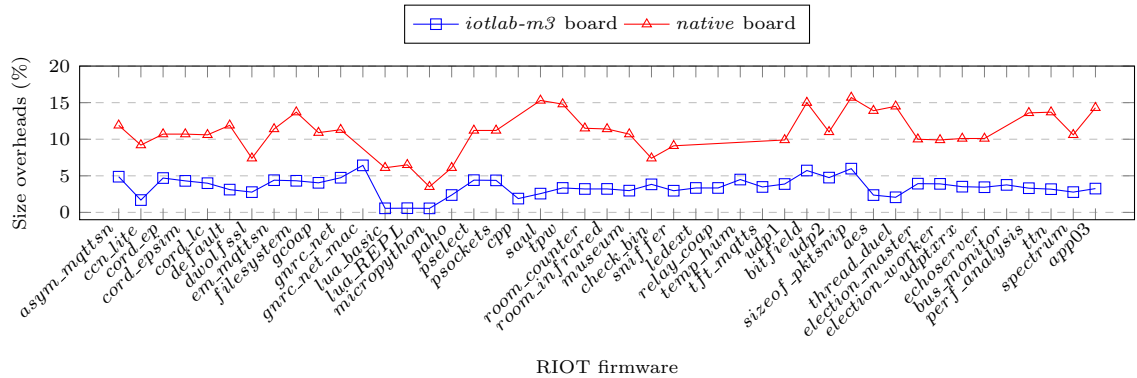

Figure \ref{fig:size-overhead} shows the percentage size overhead introduced by the protection on the firmware executable. 
The graph reflects the expected trend following the number of injected CLBs for each firmware: the average size overhead is higher in the native w.r.t. the iotlab-m3 board.
In the first case, the average size overhead is 11.3\% with a standard deviation of 2.8\%; in the latter case, it remains within the 5\% range (avg. 3.52\%, st. dev. 1.32\%).
In terms of absolute size, the actual increase consists of a few kilobytes. The native board has an average growth of 134.3KB and a standard deviation of 46.7 KB. The size of the iotlab-m3 board firmware increased to 82.9KB on average, with a standard deviation of 34.4 KB.
Also, it is worth noting that the size overhead is always less than 15.8\%, even when \toolname{} injects more than 60 CLBs, thereby corresponding to a maximum growth of 197.9KB for the native board and 160.3KB for the iotlab-m3.

Then, we empirically tested the protected firmware at runtime to verify that the introduced protections did not harm the normal functioning of the software bundle. 

In detail, the runtime evaluation consisted of 5 test runs for each protected firmware; each app has been stimulated for 2 minutes with a sequence of manual inputs obtained from the \texttt{help} command of the app or extracted from the documentation (where available).
Including the iotlab-m3 physical board (among the most used in the FIT IoT-LAB testbed) in the experimental campaign allowed us to demonstrate the solution's applicability in real-world environments. 
The experimental evaluation reported that all the protected firmware samples executed correctly, i.e., they did not crash nor trigger exceptions.

\subsection{Protection Overhead}

This set of experiments aims to evaluate the runtime overhead introduced by the CLBs in a real IoT device.
To do so, we leveraged the consumption monitoring tool~\cite{consumption_monitoring} provided by IoT-LAB.
It allows to measure the energy consumption of a node in terms of current (Ampere), voltage (Volt), and power (Watt), through an INA226 hardware component.
INA226 has a programmable conversion times (CT), which
allows it to be configured to optimize the available timing requirements in a given application. 
Along with the CT, the averaging mode (AV) allows the INA226 to be more effective in reducing the noise component during the measurement.
Thus, the periodic measure (PM) (or sampling period) is given by the formula
$$ PM = CT * AV * 2 $$
We created a monitor profile with a CT of 8244µs and AV of 4, resulting in a sampling rate of almost 66ms. 


We executed both the original and the protected version of the 45 firmwares in an iotlab-m3 node with the consumption monitoring enable.
It is worth noticing that each firmware is tested under the same input commands sequence. 

\begin{figure}[h]
    \centering
    \begin{tikzpicture}
        \begin{axis}[
            xbar,
            bar width=10pt,
            enlarge x limits={abs=0.5},
            xmin=0, xmax=13.5,
            axis lines=left,
            enlarge y limits=0.3,
            enlarge x limits=0,
            ytick={1,2,3},
            yticklabels={${Current~(cA)}$,${Voltage~(V)}$,${Power~(cW)}$},
            width=0.88\textwidth, height=0.32\textwidth,
            legend style={at={(0.5,1.13)},
                    anchor=north,legend columns=-1},
            text=red
        ]

            \addplot+[mark options={black, scale=0.75},
              error bars/.cd, 
                x dir=both, 
                x explicit
            ] table [x=y, y=x,x error=error, col sep=comma] {
                x,  y,       error
                1,  2.75,    0.80
                2,  3.30,    0.0070
                3,  9.13,    2.66
            };

            \addplot+[mark options={black, scale=0.75},
              error bars/.cd, 
                x dir=both, 
                x explicit
            ] table [x=y, y=x,x error=error, col sep=comma] {
                x,  y,       error
                1,  2.80,    0.82
                2,  3.31,    0.0071
                3,  9.24,    2.80
            };

           \addlegendentry{original firmwares}
           \addlegendentry{portected firmwares}
        \end{axis}
    \end{tikzpicture}
    
    \caption{Average and standard deviation of the current, voltage, and power usage for the original and protected firmware samples.}
    \label{fig:performace-overhead}
\end{figure}
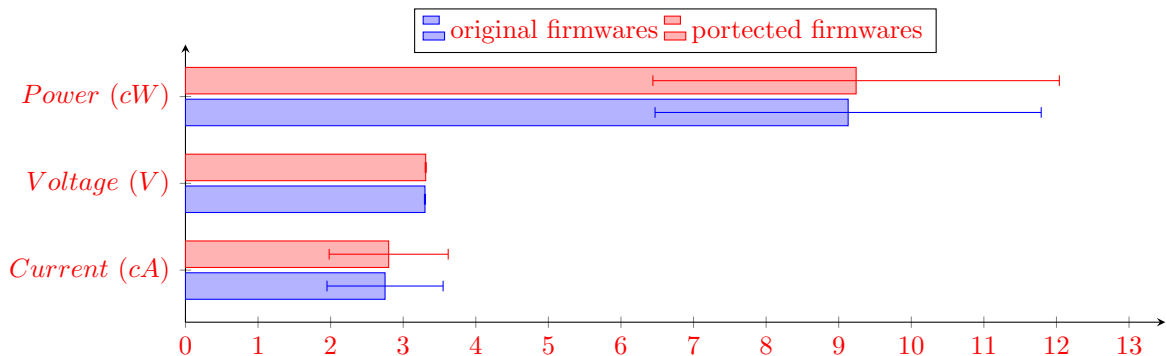

Figure~\ref{fig:performace-overhead} shows the overall mean and standard deviation for power, voltage, and current consumptions for both the original and the protected firmware.

The results highlight that the introduced protections do not significantly impact resource usage, even if the CLBs are decrypted during the firmware's executions.
In particular, the voltage and current consumptions are comparable between the original and the protected versions, which on average are 3.30 (st. dev. 0.0070) vs. 3.31 (st. dev. 0.0071) for the voltage and 2.75 (st. dev. 0.8) vs. 2.8 (st. dev. 0.82) for the current user.
However, we noticed a clear but still limited difference in the power consumption.
The protected firmware consumes on average 1.2mW more compared to their respective originals.

\subsection{Efficacy of the Protection Scheme}
\begin{figure}[h!]
    \centering
    \subcaptionbox{Execution of the original firmware. \label{fig:riot_repack}}{
        \includegraphics[width=0.64\linewidth]{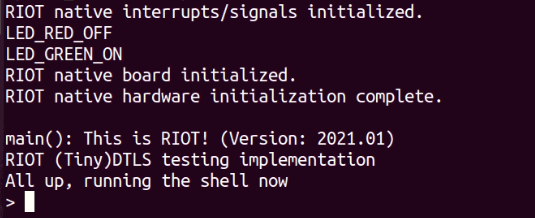}
    }%
    \hfill

    \subcaptionbox{Execution of the repackaged firmware without protection.}{
        \includegraphics[width=0.64\linewidth]{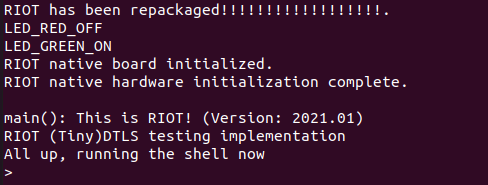}
    }%
    \hfill
    
    \subcaptionbox{Execution of the repackaged firmware protected with \toolname{}. \label{fig:riot_repack_detected}}{
        \includegraphics[width=0.64\linewidth]{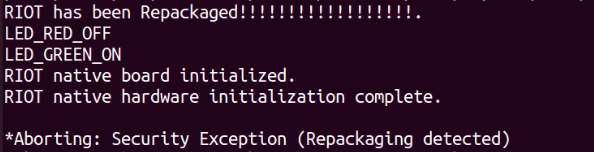}
    }%
    
    \caption{Steps of the security evaluation of \toolname{} on a firmware sample.}
    \label{fig:iot_repack_full}
\end{figure}

The last set of experiments aims to evaluate the efficacy of the protections introduced by \methodname{}.
To do so, we automatically exploited a real repackaging attack on each of the firmware of the dataset. In detail, we built a script that modifies a set of random values in the compiled firmware. 
For instance, in Figure \ref{fig:iot_repack_full}, the repackaging script replaced the original string (i.e., \texttt{RIOT native interrupts/signals initialized}) with a custom one (i.e., \texttt{RIOT has been repackaged!}). 

We repackaged all the protected firmware samples (i.e., 45 executables) and repeated the execution with the same set of inputs of the previous phase on an iotlab-m3 board hosted in the FIT IoT-LAB datacenter in Lille.
The experiments showed that 69\% of the firmware samples (i.e., 31/45) successfully detected the repackaging attempt. 
In particular:
\begin{itemize}
    \item 23 apps detect the repackaging at the startup;
    \item 8 apps detect the repackaging during the execution of the user inputs, i.e., when a specific input is sent to the firmware;
    \item 14 apps do not detect the repackaging within the end of the test. 
\end{itemize}

The results suggest a high repackaging detection rate. 
However, 31\% of the tested apps failed to detect firmware modification.
In the previous phase, we highlighted how 12 firmware images of the dataset contain only a few CLBs (i.e., less than 5). 
In fact, most of the failed detections are related to basic firmware samples that include an app with few lines of code.
For instance, Listing~\ref{lst:elementary_app_code} shows the \texttt{main} function of the \textit{twp} RIOT app.

\begin{lstlisting}[caption={Example of elementary app.}, label={lst:elementary_app_code}]
    int main(void) {
        xtimer_ticks32_t last_wakeup = xtimer_now();
        while(1) {
            xtimer_periodic_wakeup(&last_wakeup, INTERVAL);
            printf("slept until %" PRIu32 "\n", xtimer_usec_from_ticks(xtimer_now()));
        }
        return 0;
    }
\end{lstlisting}

The \texttt{main} function and the invoked ones (i.e., \texttt{xtimer\_now}, \texttt{printf}, and \texttt{xtimer\_usec\_from\_ticks}) do not contain any QC or complex statements, resulting in a lack of CLBs injected in the app code. 
In addition, the simplicity of the app's logic limited the range of potential inputs that can trigger the other CLBs included in the OS kernel, resulting in a lack of detection.
Nonetheless, for the sake of this paper, our experimental setup has been sufficient to prove that the protection scheme is reliable.
Such a consideration is supported by positive detection in the case of firmware samples with real-world apps.

\subsection{Prototype Limitations}

The experimental campaign identified some limitations of \toolname{} that will be discussed below.
In the current implementation, \moduleone{} injects the functions to decrypt and execute $ext\_fun$ as static methods in each file that contains at least a CLB.
Such a choice removes any dependency from third-party libraries that may not be natively supported by the firmware (e.g., AES) or included in the firmware bundle at the cost of introducing potentially redundant code. Moreover, this technique allows using decryption and hashing functions independently from the part of the code that contains the CLB, which may not have access to the corresponding decryption library (e.g., a CLB in the startup code).
\toolname{} implements a xor-based cipher algorithm to encrypt the CLBs to guarantee that the encrypted payload would have the same size/offset, thus avoiding potential misalignment or overrides during the binary rewriting process.

The experimental results also highlighted a direct correlation between the complexity of the $const$ value used as the encryption key and the hiding capability offered by the CLBs. 
For instance, we discovered that many constant values in the RIOT OS are 2 bytes, thus limiting the resiliency of \toolname{}  against brute force attacks to guess the encryption key and bypass the CLBs. To overcome this limitation, \toolname{} could be extended to detect different types of logic bombs, such as QCs containing a string comparison function like \texttt{strcmp()} and \texttt{strncmp()}. 

Also, it is worth noticing that even if \methodname{} supports generic firmware images, either packed in a single executable or a package (e.g., OpenWRT \cite{holt2014openwrt}), we focused our work on firmware bundles assembled as an executable file, such as the RIOT firmware. Thus, \toolname{} implements AT controls to verify a set of properties of the executable file, albeit they can be extended to check the integrity of other executables or non-code files (e.g., a configuration file).

Finally, it is worth noting that each processor provides a different set of capabilities. For instance, in the native board, we can leverage the memory management APIs of the host processor (e.g., mprotect \cite{mprotect}) to alter the memory protection bits and modify the content of the executable section (\texttt{.text}). In the case of the ARM Cortex CPU and the iotlab-m3 board, these APIs are not available, thereby requiring a change in the implementation of the decryption logic (e.g., execute/modify some methods on the \texttt{.data} section). In other words, the \methodname{} methodology can be applied to generic IoT devices, but its implementation (i.e., \toolname{}) may require slight changes depending on the target board and hardware/software constraints.

\section{Conclusion}
\label{sec:conclusion}

In this work, we proposed \methodname{}, a self-protection mechanism that ensures the resiliency of IoT firmware images against repackaging attacks through the entire production and delivery process. 
The methodology exploits the use of CLBs to hide anti-tampering checks in the firmware executable that will trigger (i.e., explode) in case of any tampering is detected.

Furthermore, we implemented \methodname{} in a tool for protecting C/C++ firmware, called \toolname{}, that is publicly available on GitHub \cite{patriot}.
\toolname{}{} modifies the firmware source code to inject CLBs and AT controls and the compiled binary to build the CLBs by encrypting specific binary portions.
It is worth emphasizing that the integrity controls do not rely on external trust anchors or verification processes.
The evaluation of 50 firmware samples for RIOT OS on two boards (i.e., \textit{native} and \textit{iotlab-m3}) demonstrated the applicability and efficacy of the tool and the proposed protection scheme at the cost of a reasonable size overhead of the firmware image.

As future extensions of this work, we plan to extend the protection scheme by adding multi-patter (i.e., heterogeneous) AT controls and evaluate the impact of the \methodname{} protection scheme on the repackaging attack steps presented in Figure~\ref{fig:repack-steps-at}.
Also, we would like to measure the computational and energy footprint of the protection scheme on resource-constrained IoT devices and extend the support to other IoT OSes, e.g., Contiki-NG and FreeRTOS.

\bibliographystyle{elsarticle-num}
\bibliography{biblio.bib}

\appendix
\section{Experimental Dataset}
\label{app:dataset}

The dataset used for evaluating \toolname{} consists of 50 firmware images built using RIOT OS 2021.05 and a different user app that has a publicly-available source repository on GitHub. 
The dataset samples use more than 130 different RIOT modules, and each app includes, on average, 12 modules with a standard deviation of 6.

Table~\ref{tbl:riotapps} reports the name of each app, the link to the GitHub project, and the subfolder that contains the source code.
Table~\ref{tbl:app2module} details the list of all the RIOT OS modules included for each firmware in the dataset.
We also specified \texttt{\$BOARD\$} and \texttt{\$DRIVER\$} to collectively refer to the additional modules needed by a firmware to execute in a particular board/driver.

\begin{table}[]
\footnotesize
\centering
\begin{tabular}{l|l|l}
    \toprule

    \textbf{App Name} &
    \textbf{GitHub Repo} &
    \textbf{Folder} 
    \\ \midrule

asym\_mqttsn & {RIOT-OS/RIOT} & examples/asymcute\_mqttsn \\ 
ccn\_lite & {RIOT-OS/RIOT} & examples/ccn-lite-relay \\ 
cord\_ep & {RIOT-OS/RIOT} & examples/cord\_ep \\ 
cord\_epsim & {RIOT-OS/RIOT} & examples/cord\_epsim \\ 
cord\_lc & {RIOT-OS/RIOT} & examples/cord\_lc \\ 
default & {RIOT-OS/RIOT} & examples/default \\ 
decho & {RIOT-OS/RIOT} & examples/dtls-echo \\ 
dsock & {RIOT-OS/RIOT} & examples/dtls-sock \\ 
dwolfssl & {RIOT-OS/RIOT} & examples/dtls-wolfssl \\ 
em\_mqttsn & {RIOT-OS/RIOT} & examples/emcute\_mqttsn \\ 
filesystem & {RIOT-OS/RIOT} & examples/filesystem \\ 
gcoap & {RIOT-OS/RIOT} & examples/gcoap \\ 
gnrc\_brouter & {RIOT-OS/RIOT} & examples/gnrc\_border\_router \\ 
gnrc\_net & {RIOT-OS/RIOT} & examples/gnrc\_networking \\ 
gnrc\_net\_mac & {RIOT-OS/RIOT} & examples/gnrc\_networking\_mac \\ 
lua\_basic & {RIOT-OS/RIOT} & examples/lua\_basic \\ 
lua\_REPL & {RIOT-OS/RIOT} & examples/lua\_REPL \\ 
micropython & {RIOT-OS/RIOT} & examples/micropython \\ 
nanocoap & {RIOT-OS/RIOT} & examples/nanocoap\_server \\ 
ndn & {RIOT-OS/RIOT} & examples/ndn-ping \\ 
paho & {RIOT-OS/RIOT} & examples/paho-mqtt \\ 
pselect & {RIOT-OS/RIOT} & examples/posix\_select \\ 
psockets & {RIOT-OS/RIOT} & examples/posix\_sockets \\ 
cpp & {RIOT-OS/RIOT} & examples/riot\_and\_cpp \\ 
saul & {RIOT-OS/RIOT} & examples/saul \\ 
tpw & {RIOT-OS/RIOT} & examples/timer\_periodic\_wakeup \\ 
room\_counter & {ARte-team/ARte} & src/Boards/room\_counter\_emcute \\ 
room\_infrared & {ARte-team/ARte} & src/Boards/room\_infrared\_emcute \\ 
museum & {ARte-team/ARte} & src/Boards/museum\_counter\_emcute \\ 
check\_bin & {andreamazzitelli/checkBin} & RiotCode \\ 
sniffer & {fu-ilab-swp2021/LoRa-Packet-Sniffer} & b-l072z \\ 
ledext & {ichatz/riotos-apps} & ledext \\ 
relay\_coap & {ichatz/riotos-apps} & relay\_coap \\ 
temp\_hum & {ichatz/riotos-apps} & temperature\_humidity \\ 
tft\_mqtts & {ichatz/riotos-apps} & tft\_mqtts \\ 
udp1 & {ichatz/riotos-apps} & udp\_usb \\ 
bitfield & {miri64/RIOT\_playground} & bitfield\_test/bitfield \\ 
udp2 & {miri64/RIOT\_playground} & udp\_test \\ 
sizeof\_pktsnip & {miri64/RIOT\_playground} & sizeof\_pktsnip \\ 
aes & {deus778/riot-aes-benchmark} &  \\ 
thread-duel & {kfessel/riot-thread-duel} &  \\ 
election\_master & {maconard/RIOT\_leader-election} & cpsiot\_masternode \\ 
election\_worker & {maconard/RIOT\_leader-election} & cpsiot\_workernode \\ 
udptxrx & {induarun9086/RIOT\_UDP\_EchoServerExample} & udptxrx \\ 
echoserver & {induarun9086/RIOT\_UDP\_EchoServerExample} & udpechoserver \\ 
bus\_monitor & {FrancescoCrino/ConnectedBusMonitor} & src/proto\_ethos \\ 
perf\_analysis & {StefanoMilani/RIOT-OS-examples} & performance-analysis \\ 
ttn & {yegorich/ttn-mapper-riot} &  \\ 
spectrum & {RIOT-OS/applications} & spectrum-scanner \\ 
app03 & {Ciusss89/gtip-riotos} & test\_03 \\ 
\bottomrule
\end{tabular}
\caption{List of RIOT apps composing the dataset.}
\label{tbl:riotapps}
\end{table}

\newpage

\begin{footnotesize}
\centering
\begin{longtable}{l|l}
\caption{RIOT OS Modules included in the firmware images of the dataset.}
\label{tbl:app2module} \\
    \toprule

    \textbf{Firmware Name} &
    \textbf{Included RIOT Modules}
    \\ \midrule

\texttt{asym\_mqtt} & \begin{tabular}[x]{@{}l@{}}\texttt{\$BOARD\$}, \texttt{asymcute}, \texttt{auto\_init}, \texttt{auto\_init\_gnrc\_netif}, \texttt{gnrc\_icmpv6\_echo}, \texttt{gnrc\_ipv6\_default},\\\texttt{netdev\_default}, \texttt{ps}, \texttt{shell}, \texttt{shell\_cmds\_default}\end{tabular} \\ \midrule[0.1mm]
\texttt{ccn\_lite} & \begin{tabular}[x]{@{}l@{}}\texttt{\$BOARD\$}, \texttt{auto\_init}, \texttt{auto\_init\_gnrc\_netif}, \texttt{ccn-lite}, \texttt{gnrc\_pktdump}, \texttt{netdev\_default},\\\texttt{prng\_xorshift}, \texttt{ps}, \texttt{shell}, \texttt{shell\_cmds\_default}\end{tabular} \\ \midrule[0.1mm]
\texttt{cord\_ep} & \begin{tabular}[x]{@{}l@{}}\texttt{\$BOARD\$}, \texttt{auto\_init}, \texttt{auto\_init\_gnrc\_netif}, \texttt{cord\_ep\_standalone}, \texttt{fmt}, \texttt{gnrc\_icmpv6\_echo},\\\texttt{gnrc\_ipv6\_default}, \texttt{netdev\_default}, \texttt{ps}, \texttt{shell}, \texttt{shell\_cmds\_default}\end{tabular} \\ \midrule[0.1mm]
\texttt{cord\_epsim} & \begin{tabular}[x]{@{}l@{}}\texttt{\$BOARD\$}, \texttt{auto\_init}, \texttt{auto\_init\_gnrc\_netif}, \texttt{cord\_epsim}, \texttt{gnrc\_ipv6\_default}, \texttt{netdev\_default},\\\texttt{xtimer}\end{tabular} \\ \midrule[0.1mm]
\texttt{cord\_lc} & \begin{tabular}[x]{@{}l@{}}\texttt{\$BOARD\$}, \texttt{auto\_init}, \texttt{auto\_init\_gnrc\_netif}, \texttt{cord\_lc}, \texttt{gnrc\_icmpv6\_echo}, \texttt{gnrc\_ipv6\_default},\\\texttt{netdev\_default}, \texttt{ps}, \texttt{shell}, \texttt{shell\_cmds\_default}\end{tabular} \\ \midrule[0.1mm]
\texttt{default} & \begin{tabular}[x]{@{}l@{}}\texttt{\$BOARD\$}, \texttt{auto\_init}, \texttt{auto\_init\_gnrc\_netif}, \texttt{gnrc}, \texttt{gnrc\_pktdump}, \texttt{gnrc\_txtsnd},\\\texttt{mci}, \texttt{netdev\_default}, \texttt{ps}, \texttt{random}, \texttt{saul\_default}, \texttt{schedstatistics},\\\texttt{shell}, \texttt{shell\_cmds\_default}\end{tabular} \\ \midrule[0.1mm]
\texttt{decho} & \begin{tabular}[x]{@{}l@{}}\texttt{\$BOARD\$}, \texttt{auto\_init}, \texttt{auto\_init\_gnrc\_netif}, \texttt{gnrc\_ipv6\_default}, \texttt{netdev\_default}, \texttt{prng\_sha1prng},\\\texttt{shell}, \texttt{shell\_cmds\_default}, \texttt{sock\_udp}, \texttt{tinydtls}\end{tabular} \\ \midrule[0.1mm]
\texttt{dsock} & \begin{tabular}[x]{@{}l@{}}\texttt{\$BOARD\$}, \texttt{auto\_init}, \texttt{auto\_init\_gnrc\_neti}, \texttt{gnrc\_ipv6\_default}, \texttt{netdev\_default}, \texttt{prng\_sha1prng},\\\texttt{shell}, \texttt{shell\_cmds\_default}, \texttt{sock\_dtls}, \texttt{sock\_udp}, \texttt{sock\_util}, \texttt{tinydtls}\end{tabular} \\ \midrule[0.1mm]
\texttt{dwolfssl} & \begin{tabular}[x]{@{}l@{}}\texttt{\$BOARD\$}, \texttt{auto\_init}, \texttt{auto\_init\_gnrc\_netif}, \texttt{gnrc\_ipv6\_default}, \texttt{netdev\_default}, \texttt{shell},\\\texttt{shell\_cmds\_default}, \texttt{sock\_udp}, \texttt{wolfcrypt}, \texttt{wolfcrypt\_dh}, \texttt{wolfcrypt\_ecc}, \texttt{wolfcrypt\_rsa},\\\texttt{wolfssl}, \texttt{wolfssl\_dtls}, \texttt{wolfssl\_psk}\end{tabular} \\ \midrule[0.1mm]
\texttt{em\_mqtt} & \begin{tabular}[x]{@{}l@{}}\texttt{\$BOARD\$}, \texttt{auto\_init}, \texttt{auto\_init\_gnrc\_netif}, \texttt{emcute}, \texttt{gnrc\_icmpv6\_echo}, \texttt{gnrc\_ipv6\_default},\\\texttt{gnrc\_netif\_single}, \texttt{netdev\_default}, \texttt{ps}, \texttt{shell}, \texttt{shell\_cmds\_default}\end{tabular} \\ \midrule[0.1mm]
\texttt{filesystem} & \begin{tabular}[x]{@{}l@{}}\texttt{\$BOARD\$}, \texttt{auto\_init}, \texttt{constfs}, \texttt{devfs}, \texttt{ps}, \texttt{shell},\\\texttt{shell\_cmds\_default}, \texttt{vfs\_auto\_format}, \texttt{vfs\_default}\end{tabular} \\ \midrule[0.1mm]
\texttt{gcoap} & \begin{tabular}[x]{@{}l@{}}\texttt{\$BOARD\$}, \texttt{auto\_init}, \texttt{auto\_init\_gnrc\_netif}, \texttt{fmt}, \texttt{gcoap}, \texttt{gnrc\_icmpv6\_echo},\\\texttt{gnrc\_ipv6\_default}, \texttt{ipv4\_addr}, \texttt{ipv6\_addr}, \texttt{lwip\_arp}, \texttt{lwip\_dhcp\_auto}, \texttt{lwip\_ipv4},\\\texttt{lwip\_ipv6}, \texttt{lwip\_ipv6\_autoconfig}, \texttt{lwip\_netdev}, \texttt{netdev\_default}, \texttt{netutils}, \texttt{od},\\\texttt{ps}, \texttt{random}, \texttt{shell}, \texttt{shell\_cmds\_default}, \texttt{socket\_zep}\end{tabular} \\ \midrule[0.1mm]
\texttt{gnrc\_brouter} & \begin{tabular}[x]{@{}l@{}}\texttt{\$BOARD\$}, \texttt{auto\_init}, \texttt{auto\_init\_gnrc\_netif}, \texttt{gnrc\_dhcpv6\_client\_6lbr}, \texttt{gnrc\_icmpv6\_echo}, \\\texttt{gnrc\_ipv6\_auto\_subnets\_simple},\texttt{gnrc\_ipv6\_nib\_dns}, \texttt{gnrc\_rpl}, \texttt{gnrc\_sixlowpan\_border\_router\_default},\\ \texttt{gnrc\_uhcpc}, \texttt{netdev\_default}, \texttt{ps}, \texttt{shell}, \texttt{shell\_cmds\_default}, \texttt{sock\_dns}\end{tabular} \\ \midrule[0.1mm]
\texttt{gnrc\_net} & \begin{tabular}[x]{@{}l@{}}\texttt{\$BOARD\$}, \texttt{auto\_init}, \texttt{auto\_init\_gnrc\_netif}, \texttt{auto\_init\_gnrc\_rpl}, \texttt{gnrc\_icmpv6\_echo}, \texttt{gnrc\_icmpv6\_error},\\\texttt{gnrc\_ipv6\_nib\_dns}, \texttt{gnrc\_ipv6\_router\_default}, \texttt{gnrc\_rpl}, \texttt{netdev\_default}, \texttt{netstats\_ipv6}, \texttt{netstats\_l2},\\\texttt{netstats\_rpl}, \texttt{ps}, \texttt{shell}, \texttt{shell\_cmd\_gnrc\_udp}, \texttt{shell\_cmds\_default}, \texttt{sock\_dns}, \texttt{socket\_zep}\end{tabular} \\ \midrule[0.1mm]
\texttt{gnrc\_net\_mac} & \begin{tabular}[x]{@{}l@{}}\texttt{\$BOARD\$}, \texttt{auto\_init}, \texttt{auto\_init\_gnrc\_netif}, \texttt{auto\_init\_gnrc\_rpl}, \texttt{gnrc\_gomach}, \texttt{gnrc\_icmpv6\_echo},\\\texttt{gnrc\_ipv6\_router\_default}, \texttt{gnrc\_lwmac}, \texttt{gnrc\_pktdump}, \texttt{gnrc\_rpl}, \texttt{gnrc\_udp}, \texttt{netdev\_default},\\\texttt{netstats\_ipv6}, \texttt{netstats\_l2}, \texttt{netstats\_rpl}, \texttt{ps}, \texttt{shell}, \texttt{shell\_cmds\_default}\end{tabular} \\ \midrule[0.1mm]
\texttt{lua\_basic} & \texttt{\$BOARD\$}, \texttt{auto\_init}, \texttt{lua}  \\ \midrule[0.1mm]
\texttt{lua\_REPL} & \texttt{\$BOARD\$}, \texttt{auto\_init}, \texttt{lua}  \\ \midrule[0.1mm]
\texttt{micropython} & \texttt{\$BOARD\$}, \texttt{auto\_init}, \texttt{micropython}  \\ \midrule[0.1mm]
\texttt{nanocoap} & \begin{tabular}[x]{@{}l@{}}\texttt{\$BOARD\$}, \texttt{auto\_init}, \texttt{auto\_init\_gnrc\_netif}, \texttt{fmt}, \texttt{gnrc\_icmpv6\_echo}, \texttt{gnrc\_ipv6\_default},\\\texttt{hashes}, \texttt{nanocoap\_sock}, \texttt{netdev\_default}, \texttt{prng\_minstd}, \texttt{sock\_udp}, \texttt{xtimer}\end{tabular} \\ \midrule[0.1mm]
\texttt{ndn} & \begin{tabular}[x]{@{}l@{}}\texttt{\$BOARD\$}, \texttt{auto\_init}, \texttt{auto\_init\_gnrc\_netif}, \texttt{ndn-riot}, \texttt{netdev\_default}, \texttt{random},\\\texttt{shell}, \texttt{shell\_cmds\_default}\end{tabular} \\ \midrule[0.1mm]
\texttt{paho} & \begin{tabular}[x]{@{}l@{}}\texttt{\$BOARD\$}, \texttt{auto\_init}, \texttt{auto\_init\_gnrc\_netif}, \texttt{gnrc\_icmpv6\_echo}, \texttt{gnrc\_icmpv6\_error}, \texttt{gnrc\_ipv6\_default},\\\texttt{ipv4\_addr}, \texttt{ipv6\_addr}, \texttt{lwip}, \texttt{lwip\_arp}, \texttt{lwip\_dhcp\_auto}, \texttt{lwip\_ipv4},\\\texttt{lwip\_ipv6\_autoconfig}, \texttt{lwip\_netdev}, \texttt{netdev\_default}, \texttt{ps}, \texttt{shell}, \texttt{shell\_cmds\_default},\\\texttt{sock\_async\_event}, \texttt{sock\_ip}, \texttt{sock\_tcp}, \texttt{sock\_udp}, \texttt{ztimer}, \texttt{ztimer\_msec}\end{tabular} \\ \midrule[0.1mm]
\texttt{pselect} & \begin{tabular}[x]{@{}l@{}}\texttt{\$BOARD\$}, \texttt{auto\_init}, \texttt{auto\_init\_gnrc\_netif}, \texttt{gnrc\_ipv6\_default}, \texttt{netdev\_default}, \texttt{posix\_inet},\\\texttt{posix\_select}, \texttt{posix\_sockets}, \texttt{sock\_udp}\end{tabular} \\ \midrule[0.1mm]
\texttt{psocket} & \begin{tabular}[x]{@{}l@{}}\texttt{\$BOARD\$}, \texttt{auto\_init}, \texttt{auto\_init\_gnrc\_netif}, \texttt{gnrc\_ipv6\_default}, \texttt{netdev\_default}, \texttt{posix\_inet},\\\texttt{posix\_sleep}, \texttt{posix\_sockets}, \texttt{ps}, \texttt{shell}, \texttt{shell\_cmds\_default}, \texttt{sock\_udp}\end{tabular} \\ \midrule[0.1mm]
\texttt{cpp} & \texttt{\$BOARD\$}, \texttt{auto\_init}, \texttt{cpp}, \texttt{libstdcpp}  \\ \midrule[0.1mm]
\texttt{saul} & \texttt{\$BOARD\$}, \texttt{auto\_init}, \texttt{ps}, \texttt{saul\_default}, \texttt{shell}, \texttt{shell\_cmds\_default}  \\ \midrule[0.1mm]
\texttt{tpw} & \texttt{\$BOARD\$}, \texttt{auto\_init}, \texttt{ztimer\_msec}  \\ \midrule[0.1mm]
\texttt{room\_counter} & \begin{tabular}[x]{@{}l@{}}\texttt{\$BOARD\$}, \texttt{auto\_init}, \texttt{auto\_init\_gnrc\_netif}, \texttt{emcute}, \texttt{gnrc\_icmpv6\_echo}, \texttt{gnrc\_ipv6\_default},\\\texttt{gnrc\_netdev\_default}, \texttt{gnrc\_sock\_udp}, \texttt{ps}, \texttt{shell}, \texttt{shell\_commands}, \texttt{xtimer}\end{tabular} \\ \midrule[0.1mm]
\texttt{room\_infrared} & \begin{tabular}[x]{@{}l@{}}\texttt{\$BOARD\$}, \texttt{auto\_init}, \texttt{auto\_init\_gnrc\_netif}, \texttt{emcute}, \texttt{gnrc\_icmpv6\_echo}, \texttt{gnrc\_ipv6\_default},\\\texttt{gnrc\_netdev\_default}, \texttt{gnrc\_sock\_udp}, \texttt{ps}, \texttt{shell}, \texttt{shell\_commands}, \texttt{xtimer}\end{tabular} \\ \midrule[0.1mm]
\texttt{museum} & \begin{tabular}[x]{@{}l@{}}\texttt{\$BOARD\$}, \texttt{auto\_init}, \texttt{auto\_init\_gnrc\_netif}, \texttt{emcute}, \texttt{gnrc\_icmpv6\_echo}, \texttt{gnrc\_ipv6\_default},\\\texttt{gnrc\_netdev\_default}, \texttt{gnrc\_sock\_udp}, \texttt{ps}, \texttt{shell}, \texttt{shell\_commands}, \texttt{xtimer}\end{tabular} \\ \midrule[0.1mm]
\texttt{check\_bin} & \begin{tabular}[x]{@{}l@{}}\texttt{\$BOARD\$}, \texttt{\$DRIVER\$}, \texttt{auto\_init}, \texttt{auto\_init\_loramac}, \texttt{fmt}, \texttt{periph\_gpio},\\\texttt{periph\_gpio\_irq}, \texttt{periph\_i2c}, \texttt{semtech-loramac}, \texttt{semtech\_loramac\_rx}, \texttt{shell}, \texttt{shell\_commands},\\\texttt{u8g2}, \texttt{xtimer}, \texttt{ztimer}, \texttt{ztimer\_usec}\end{tabular} \\ \midrule[0.1mm]
\texttt{sniffer} & \begin{tabular}[x]{@{}l@{}}\texttt{\$BOARD\$}, \texttt{\$DRIVER\$}, \texttt{auto\_init}, \texttt{fatfs\_vfs}, \texttt{mtd\_sdcard}, \texttt{periph\_gpio\_irq}, \texttt{vfs}, \texttt{xtimer}\end{tabular} \\ \midrule[0.1mm]
\texttt{ledex} & \texttt{\$BOARD\$}, \texttt{auto\_init}, \texttt{periph\_gpio}, \texttt{xtimer}  \\ \midrule[0.1mm]
\texttt{relay\_coap} & \texttt{\$BOARD\$}, \texttt{auto\_init}, \texttt{periph\_gpio}, \texttt{xtimer}  \\ \midrule[0.1mm]
\texttt{temp\_hum} & \texttt{\$BOARD\$}, \texttt{auto\_init}, \texttt{dht}, \texttt{fmt}, \texttt{periph\_rtc}  \\ \midrule[0.1mm]
\texttt{tft\_mqtt} & \begin{tabular}[x]{@{}l@{}}\texttt{\$BOARD\$}, \texttt{auto\_init}, \texttt{auto\_init\_gnrc\_netif}, \texttt{emcute}, \texttt{gnrc\_ipv6\_default}, \texttt{gnrc\_netdev\_default},\\\texttt{gnrc\_netif\_single}, \texttt{gnrc\_uhcpc}, \texttt{periph\_gpio}, \texttt{periph\_spi}, \texttt{stdio\_ethos}, \texttt{ucglib},\\\texttt{xtimer}\end{tabular} \\ \midrule[0.1mm]
\texttt{udp1} & \begin{tabular}[x]{@{}l@{}}\texttt{\$BOARD\$}, \texttt{auto\_init}, \texttt{auto\_init\_gnrc\_netif}, \texttt{gnrc\_icmpv6\_echo}, \texttt{gnrc\_ipv6\_default}, \texttt{gnrc\_netdev\_default},\\\texttt{gnrc\_uhcpc}, \texttt{stdio\_ethos}, \texttt{xtimer}\end{tabular} \\ \midrule[0.1mm]
\texttt{bitfield} & \texttt{\$BOARD\$}, \texttt{auto\_init}  \\ \midrule[0.1mm]
\texttt{udp2} & \begin{tabular}[x]{@{}l@{}}\texttt{\$BOARD\$}, \texttt{auto\_init}, \texttt{auto\_init\_gnrc\_netif}, \texttt{auto\_init\_gnrc\_rpl}, \texttt{gnrc\_icmpv6\_echo},\\\texttt{gnrc\_ipv6\_router\_default},\texttt{gnrc\_netdev\_default}, \texttt{gnrc\_rpl}, \texttt{gnrc\_udp}, \texttt{netstats\_ipv6}, \texttt{netstats\_l2}, \texttt{od},\\\texttt{ps}, \texttt{shell}, \texttt{shell\_commands}\end{tabular} \\ \midrule[0.1mm]
\texttt{sizeof\_pktsnip} & \texttt{fmt}, \texttt{\$BOARD\$}  \\ \midrule[0.1mm]
\texttt{aes} & \begin{tabular}[x]{@{}l@{}}\texttt{\$BOARD\$}, \texttt{auto\_init}, \texttt{cipher\_modes}, \texttt{crypto\_aes\_128}, \texttt{crypto\_aes\_192}, \texttt{crypto\_aes\_256},\\\texttt{od}, \texttt{od\_string}, \texttt{random}, \texttt{shell}, \texttt{shell\_commands}, \texttt{xtimer}\end{tabular} \\ \midrule[0.1mm]
\texttt{thread-duel} & \texttt{\$BOARD\$}, \texttt{auto\_init}, \texttt{sched\_cb}, \texttt{sema}, \texttt{xtimer}  \\ \midrule[0.1mm]
\texttt{election\_master} & \begin{tabular}[x]{@{}l@{}}\texttt{\$BOARD\$}, \texttt{auto\_init}, \texttt{auto\_init\_gnrc\_netif}, \texttt{auto\_init\_gnrc\_rpl}, \texttt{gnrc}, \texttt{gnrc\_icmpv6\_echo},\\\texttt{gnrc\_icmpv6\_error}, \texttt{gnrc\_ipv6\_default}, \texttt{gnrc\_ipv6\_router\_default}, \texttt{gnrc\_netdev\_default}, \texttt{gnrc\_pktdump},\\\texttt{gnrc\_rpl},\texttt{gnrc\_sock\_udp}, \texttt{gnrc\_txtsnd}, \texttt{gnrc\_udp}, \texttt{netstats\_ipv6}, \texttt{netstats\_l2},\\\texttt{netstats\_rpl},\texttt{periph\_rtc}, \texttt{ps}, \texttt{random}, \texttt{schedstatistics}, \texttt{shell}, \texttt{shell\_commands}, \texttt{xtimer}\end{tabular} \\ \midrule[0.1mm]
\texttt{election\_worker} & \begin{tabular}[x]{@{}l@{}}\texttt{\$BOARD\$}, \texttt{auto\_init}, \texttt{auto\_init\_gnrc\_netif}, \texttt{auto\_init\_gnrc\_rpl}, \texttt{gnrc}, \texttt{gnrc\_icmpv6\_echo},\\\texttt{gnrc\_icmpv6\_error}, \texttt{gnrc\_ipv6\_default}, \texttt{gnrc\_ipv6\_router\_default}, \texttt{gnrc\_netdev\_default}, \texttt{gnrc\_pktdump},\\\texttt{gnrc\_rpl},\texttt{gnrc\_sock\_udp}, \texttt{gnrc\_txtsnd}, \texttt{gnrc\_udp}, \texttt{netstats\_ipv6}, \texttt{netstats\_l2},\\\texttt{netstats\_rpl},\texttt{periph\_rtc}, \texttt{ps}, \texttt{random}, \texttt{schedstatistics}, \texttt{shell}, \texttt{shell\_commands}, \texttt{xtimer}\end{tabular} \\ \midrule[0.1mm]
\texttt{udptxrx} & \begin{tabular}[x]{@{}l@{}}\texttt{\$BOARD\$}, \texttt{auto\_init}, \texttt{auto\_init\_gnrc\_netif}, \texttt{auto\_init\_gnrc\_rpl}, \texttt{gnrc\_ipv6\_default}, \texttt{gnrc\_netdev\_default},\\\texttt{gnrc\_rpl}, \texttt{gnrc\_sock\_udp}, \texttt{gnrc\_udp}\end{tabular} \\ \midrule[0.1mm]
\texttt{echoserver} & \begin{tabular}[x]{@{}l@{}}\texttt{\$BOARD\$}, \texttt{auto\_init}, \texttt{auto\_init\_gnrc\_netif}, \texttt{auto\_init\_gnrc\_rpl}, \texttt{gnrc\_ipv6\_default}, \texttt{gnrc\_netdev\_default},\\\texttt{gnrc\_rpl}, \texttt{gnrc\_sock\_udp}, \texttt{gnrc\_udp}\end{tabular} \\ \midrule[0.1mm]
\texttt{bus\_monitor} & \begin{tabular}[x]{@{}l@{}}\texttt{\$BOARD\$}, \texttt{\$DRIVER\$}, \texttt{auto\_init}, \texttt{auto\_init\_gnrc\_netif}, \texttt{emcute}, \texttt{gnrc\_icmpv6\_echo},\\\texttt{gnrc\_ipv6\_default}, \texttt{gnrc\_netdev\_default}, \texttt{gnrc\_uhcpc}, \texttt{hts221}, \texttt{ps}, \texttt{shell},\\\texttt{shell\_commands}, \texttt{stdio\_ethos}, \texttt{xtimer}, \texttt{ztimer}, \texttt{ztimer\_msec}\end{tabular} \\ \midrule[0.1mm]
\texttt{perf\_analysis} & \begin{tabular}[x]{@{}l@{}}\texttt{\$BOARD\$}, \texttt{auto\_init}, \texttt{crypto}, \texttt{prng\_minstd}, \texttt{ps}, \texttt{shell}, \texttt{shell\_commands}, \texttt{xtimer}\end{tabular} \\ \midrule[0.1mm]
\texttt{ttn} & \texttt{\$BOARD\$}, \texttt{auto\_init}, \texttt{minmea}, \texttt{periph\_uart}, \texttt{xtimer}  \\ \midrule[0.1mm]
\texttt{spectrum} & \begin{tabular}[x]{@{}l@{}}\texttt{\$BOARD\$}, \texttt{auto\_init}, \texttt{auto\_init\_gnrc\_netif}, \texttt{fmt}, \texttt{gnrc}, \texttt{netdev\_default},\\\texttt{xtimer}, \texttt{ztimer64\_xtimer\_compat}\end{tabular} \\ \midrule[0.1mm]
\texttt{app03} & \begin{tabular}[x]{@{}l@{}}\texttt{\$BOARD\$}, \texttt{auto\_init}, \texttt{ps}, \texttt{shell}, \texttt{shell\_commands}, \texttt{uptime},\texttt{xtimer}\end{tabular} \\

\bottomrule
\end{longtable}
\end{footnotesize}

\end{document}